\documentclass[a4paper,fleqn]{cas-dc}

\usepackage[numbers]{natbib}
\usepackage{amsmath,mathtools}
\usepackage{physics}

\def\tsc#1{\csdef{#1}{\textsc{\lowercase{#1}}\xspace}}
\tsc{WGM}
\tsc{QE}
\tsc{EP}
\tsc{PMS}
\tsc{BEC}
\tsc{DE}

\begin{document}
\let\WriteBookmarks\relax
\def\floatpagepagefraction{1}
\def\textpagefraction{.001}
\shorttitle{Stochastic dynamics characterization via DMD}
\shortauthors{A. Baratz et~al.}

\title{Data-Driven Reconstruction and Characterization of Stochastic Dynamics via Dynamical Mode Decomposition
}                      



\author[1,2]{Adva Baratz}
\cormark[1]
\ead{adva.baratz@gmail.com}

\credit{Conceptualization, Investigation, Methodology, Formal Analysis, Writing - Original Draft}


\author[3]{Loris Maria Cangemi}

\credit{Investigation, Data Curation, Methodology, Writing - Original Draft}

\author[4]{Assaf Hamo}
\credit{Investigation,  Writing-review \& editing}




\author[2]{Sivan Refaely-Abramson}
\credit{Investigation,  Writing-review \& editing}


\author[5,6,7]{Amikam Levy}
\cormark[1]
\ead{amikam.levy@biu.ac.il}

\credit{Conceptualization, Supervision, Investigation, Methodology, Funding Acquisition,  Writing - Original Draft}

\affiliation[1]{Institute of Artificial Intelligence, Weizmann Institute of Science, Rehovot 7610001, Israel}
\affiliation[2]{Department of Molecular Chemistry and Materials Science, Weizmann Institute of Science, Rehovot 7610001, Israel}
\affiliation[3]{Department of Electrical Engineering and Information Technology, Università degli Studi di Napoli Federico II, via Claudio 21, Napoli, 80125, Italy}
\affiliation[4]{Department of Physics, Bar-Ilan University,Ramat-Gan 52900, Israel}
\affiliation[5]{Department of Chemistry, Bar-Ilan University, Ramat-Gan 52900, Israel}
\affiliation[6]{ Institute of Nanotechnology and Advanced Materials, Bar-Ilan University, Ramat-Gan 52900, Israel}
\affiliation[7]{Center for Quantum Entanglement Science and Technology, Bar-Ilan University, Ramat-Gan 52900, Israel}

\cortext[cor1]{Corresponding author}


\begin{abstract}
Noise fundamentally limits the performance and predictive capabilities of classical and quantum dynamical systems by degrading stability and obscuring intrinsic dynamical characteristics. Characterizing such noise accurately is essential for enhancing measurement precision, understanding environmental interactions, and designing effective control strategies across diverse scientific and engineering domains. However, extracting the environment spectral features and associated characteristic decay or coherence times from limited and noisy datasets remains challenging. Here, we introduce a general, data-driven framework based on Dynamical Mode Decomposition (DMD) to analyze system dynamics under stochastic noise. We reinterpret DMD modes as statistical weights over an ensemble of stochastic trajectories and, via a nonlinear mapping, construct a PSD-like spectral fingerprint of the noise. This enables the identification of dominant frequency contributions in both broadband (white) and correlated ($1/f$) noise environments, as well as direct extraction of intrinsic characteristic decay times from DMD eigenvalues. To overcome instability in standard DMD-based extrapolation, we develop a constrained reconstruction method using extracted decay times as physical bounds and the learned noise as spectral weights. We demonstrate the effectiveness of this approach through simulations of quantum system dynamics subject to decoherence from noise, demonstrating its robustness and predictive capabilities and comparing it with standard methods. This methodology provides a trajectory-level framework for diagnostic and predictive analysis of stochastic processes from limited, noisy time-series data.
\end{abstract}

\begin{graphicalabstract}
\end{graphicalabstract}

\begin{highlights}
\item A data-driven Dynamical Mode Decomposition (DMD) based method is developed to characterize stochastic dynamics from short, noisy trajectory ensembles without assuming a parametric noise model. 
\item DMD mode amplitudes are reinterpreted as ensemble-level statistical weights and mapped via a nonlinear transformation to produce a normalized PSD-like spectral fingerprint. 
\item 
Characteristic coherence times are extracted directly from the DMD eigenvalue spectrum, providing a model-free diagnostic of stability and decoherence.
\item A constrained reconstruction scheme stabilizes long-time DMD extrapolation using the extracted decay time and learned spectral weights.
\item The approach is validated on simulated dephasing dynamics and benchmarked against standard spectral estimators and inversion methods (Supplementary Material).
\end{highlights}


\maketitle

\section{Introduction}

Understanding, characterizing, and mitigating noise is a fundamental challenge across numerous scientific and engineering disciplines. Accurately reconstructing the spectral structure of stochastic noise is critical to identifying key dynamical parameters that dictate system stability, coherence, and controllability~\cite{bylander2011noise,norris2016qubit}. Central to noise characterization is the power spectral density (PSD), which encodes the distribution of environmental fluctuations across frequencies. White noise, characterized by frequency-independent fluctuations, and correlated noise, such as $1/f$ noise exhibiting persistent low-frequency correlations~\cite{paladino2002decoherence,bergli2009decoherence,paladino20141}, significantly affect performance across diverse applications, from quantum technologies to biological systems and financial markets~\cite{gardiner2004handbook}. Accurate PSD reconstruction is thus essential for diagnosing noise mechanisms and designing targeted control strategies~\cite{gordon2007universal,cangemi2025theory}.

Another important property directly impacted by noise is the intrinsic characteristic decay or coherence time $T_2^*$, quantifying the rate at which systems lose coherence or stability due to environmental interactions. Precise estimation of these decay times is essential in applications ranging from quantum computing, sensing, and spectroscopy to mechanical systems and biological processes. However, reliably determining these time scales from limited and noisy experimental data is challenging: measurement noise, readout instability, and limited sampling obscure the coherence envelope. Traditional approaches such as Fourier or Welch analysis~\cite{Welch1967} typically require long, high-fidelity datasets, while Bayesian inference demands strong prior models. Even advanced harmonic inversion methods like Filter-Diagonalization~\cite{neuhauser1990bound,wall1995extraction,mandelshtam2001fdm,cohn2021quantum} are limited by short or broadband stochastic data.

Recently, data-driven methods and machine learning (ML) models have emerged as promising tools for denoising and learning the underlying quantum dynamics from noisy data~\cite{Dunjko_2018,PhysRevA.107.010101,PhysRevLett.132.100602,Ronellenfitsch2018}. These include supervised models such as neural networks~\cite{LeCun2015,BruntonKutz2019}, unsupervised approaches like principal component analysis~\cite{Jolliffe2002,JOSSE20121869}, and time-series models such as recurrent neural networks~\cite{LeCun2015,bengio1994learning,elman1990finding} and long short-term memory networks~\cite{hochreiter1997long}. However, many of these approaches require extensive training or lack physical interpretability, particularly in systems with strong transients or non-stationary noise.

Another unsupervised time-series method is Dynamical Mode Decomposition (DMD), which models a system’s evolution by decomposing complex dynamics into spatial-temporal modes~\cite{schmid2010dynamic,Tu2014}. Although originally developed in fluid mechanics, where the dynamics are governed by the Navier–Stokes equations, DMD is a data-driven approach that operates directly on measurements and does not require prior knowledge of the underlying equations. As such, it has been applied across a wide range of disciplines involving complex dynamical behavior, well beyond the scope of fluid-like or Navier–Stokes-type systems \cite{kutz2016dynamic,BRUNTON20161_EEG,DMD_Infectious_Desies,PhysRevResearch_DMD,DMD_Power_Grid}. A key advantage of DMD is its ability not only to analyze complex dynamics from limited datasets but also to extrapolate system evolution beyond the measurement window. This is desirable across many scientific fields where the acquisition of long-duration measurements is constrained by low signal-to-noise ratios or data storage limitations.
However, while DMD provides physically interpretable insights into system behavior, its standard formulation struggles to decompose dynamics in the presence of non-stationary and strong transients~\cite{Rowley2009SpectralAO,proctor2016dynamic,Schmid2022DMDReview,wu2021challenges}. This limits its applicability in realistic scenarios where the underlying system is subject to stochastic noise. Moreover, stochastic fluctuations introduce significant instability into the coherence dynamics decomposed by DMD within the measurement window, fundamentally compromising its ability to reliably extrapolate future evolution.

Here, we propose an extension of the DMD framework for analyzing stochastic trajectory ensembles directly from measured time traces, without assuming a parametric model for the underlying noise. Although we validate the method using qubit dephasing dynamics, the framework is formulated at the trajectory level and is therefore not tied to this specific quantum model; similar stochastic phase-accumulation structure also appears in classical phase-noise problems, such as oscillators with fluctuating instantaneous frequency.

We construct a reduced DMD space that not only captures the system's essential evolution but also infers key physical properties. We interpret DMD's spatial-temporal modes as statistical weights over stochastic realizations and use them to construct a normalized PSD-like spectrum, or more precisely, a normalized spectral-weight distribution of the noise. We further demonstrate that $T_2^*$ can be directly extracted from the DMD spectrum and used as a physical constraint to suppress unphysical long-time growth in DMD extrapolation. With these components in place, we introduce a novel reconstruction formula that replaces the standard DMD weighting scheme with the learned noise spectrum and applies eigenvalue constraints based on $T_2^*$, enabling prediction of the system's evolution beyond the original measurement window.

\begin{figure}
\includegraphics[width=0.8\linewidth]{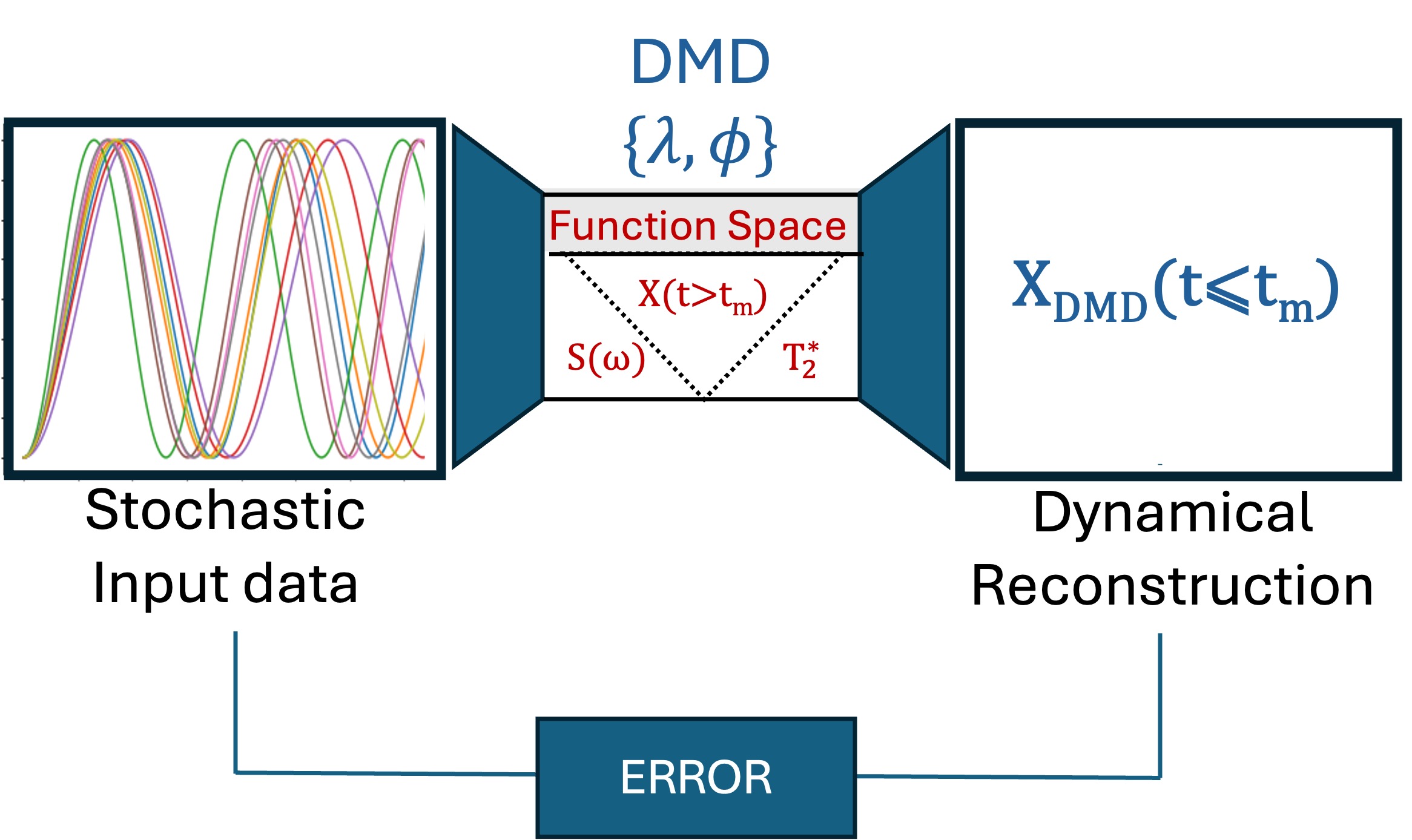} 
\caption{A schematic illustration of our data-driven approach shows how stochastic input data is transformed into physically informative properties. An ensemble of realizations measured over $0\leq t \leq t_m $ is mapped by DMD into a reduced set of spatial-temporal modes, which reconstruct the original data with minimal error. From this, we extract three key physical properties: the noise spectral weights $(\text{S}(\omega))$, coherence decay $(T_2^*)$, and extrapolated dynamics beyond the measurement window $(X(t>t_m))$.}
\label{fig:Fig-1}
\end{figure}

 We demonstrate our approach through data obtained from simulations of quantum system dynamics subject to decoherence induced by both white broadband and $1/f$ correlated noise,  and include a comparison with standard methods in the Supplementary Material. Through Ramsey-type simulations of a qubit, we show that the method:
(i) reconstructs the underlying spectrum weights with minimal data, (ii) identifies $T_2^*$ as an emergent spectral feature, and (iii) enables robust prediction of long-time dynamics. These objectives are illustrated schematically in Fig.~\ref{fig:Fig-1}, which shows a diagrammatic flow from an ensemble of stochastic realizations (left) into a reduced DMD space (middle), and onward to reconstructed dynamics (right), with an error estimation stage quantifying the discrepancy between input and reconstructed data. Our method is fully data-driven, physically interpretable, and does not require training. 
 Since the analysis acts on stochastic trajectory ensembles, it provides a framework for noise characterization in quantum systems and, more broadly, in stochastic dynamical systems with comparable trajectory-level structure.

\section{
Simulation Models: Quantum Systems as an Illustrative Example}

To demonstrate our method in a controlled setting, we apply it to quantum trajectory data generated from simulations of a qubit experiencing dephasing noise, considering two representative noise scenarios. 
The first involves $1/f$ noise, generated using an ensemble of two-state fluctuators (TSFs) that stochastically switch between states following a random telegraph process. When the switching rates are drawn from a broad distribution, the cumulative spectral density of the noise follows a $1/f$ profile, which is common in many solid-state quantum devices. This model captures both Gaussian and non-Gaussian dephasing behavior, with the latter emerging when a few strongly coupled fluctuators dominate (see App.~\ref{app:1_over_f}).

In addition to $1/f$ noise, we simulate white noise using a standard Wiener process (see App.~\ref{app:white_noise}). This allows us to compare how the method behaves under uncorrelated and correlated noise conditions, helping to assess its robustness across a range of spectral structures.

The qubit is initialized in a superposition state $\frac{1}{\sqrt{2}}\left(\ket{0} - \ket{1}\right)$ and evolves under the stochastic Hamiltonian:
\begin{equation}
    H(t) = H_0 - \frac{\xi(t)}{2} \sigma_z,
    \label{eq:Hamiltonian}
\end{equation}
where $H_0 = \frac{\omega_0}{2} \sigma_z$ governs the coherent evolution and $\xi(t)$ is a time-dependent noise term. For the $1/f$ case, $\xi(t) = \sum_i \xi_i(t)$ represents a sum over 500 independent TSFs. This drives pure dephasing, with coherence decaying over a timescale $T_2^*$.

These models are broadly applicable to a range of physical qubit systems, including superconducting phase, flux, and charge qubits, as well as spin-based platforms. As a motivating experimental platform, we consider an NV$^-$ center in diamond, where the qubit is defined using the $\ket{m_s = 0}$ and $\ket{m_s = 1}$ ground states. We set $f_0=\frac{\omega_0}{2 \pi} = 1$ MHz to simulate realistic conditions, leading to a coherence time of approximately $T_2^* \sim 1~\mu$s. Full details of the simulation procedure are provided in the Materials and Methods section.
 
\section{Standard DMD}

To uncover the dynamical properties embedded in this data set, we apply DMD, a method designed to extract spatial-temporal modes and their associated dynamics from time-resolved data by determining the eigenvalues $\lambda_i$ and eigenfunctions, commonly referred to as DMD modes, $\phi_i$, of a linear propagator $\tilde{A}$. The input data for DMD consists of two matrices, $X$ and $X'$, each with dimensions $n\times (m-1)$, where $n$ is the number of realizations and $m$ is the number of time steps. Each row of $X$ represents a single realization evolving over the successive time steps ${t_{0}, \ldots, t_{m-2}}$. The rows of $X'$ represent the same realizations, shifted forward by one time step, covering ${t_{1}, \ldots,t_{m-1}}$. These matrices are linked through $\tilde{A}$, an $n\times n$ operator that approximately satisfies the relation  $X'\approx\tilde{A}X$. Using singular value decomposition (SVD), the dimensionality of $\tilde{A}$ is reduced to facilitate the computation of its eigen-decomposition. This yields the reduced operator $A_r$, where $r$ is the rank of the reduced space. Its eigenvalues $\lambda_i$ and corresponding DMD modes $\phi_i$ characterize the system's temporal and spatial behavior (see App.~\ref{app:DMD}). 

The DMD reconstruction formula, given by Eq.~(\ref{eq:DMD_reconstruct}) uses the modes $\phi_i$ and their associated eigenvalues to recover the system's temporal evolution:

\begin{equation}
    X_{\rm DMD} \approx \sum_{i=1}^{r} \phi_i b_i e^{\mu_i T},
    \label{eq:DMD_reconstruct}
\end{equation}
where $X_{\textit{\rm DMD}}$ is the reconstructed data matrix of dimensions $n\times m$, $r$ is the DMD rank, $\phi_i$ is the $i$-th DMD spatial mode, and $b_i$ is the initial amplitude of mode $i$ (see App.~\ref{app:DMD}).  
The term $e^{\mu_i T}$ represents the temporal evolution of mode $i$ over time, where $T$ is a diagonal matrix of time instants $T=diag(t_0,t_1, \ldots,t_m)$. In this formulation, $\mu_i$ is defined as the continuous-time DMD eigenvalue and is related to the discrete-time DMD eigenvalue $\lambda_i$ by:
\begin{equation}
    \mu_i = \frac{\log(\lambda_i)}{\Delta t}
\label{eq-omega}
\end{equation}
Dividing the angular frequency $\omega_i= \text{Im} (\mu_i)$ by $2\pi$ yields the corresponding natural frequency $f_i$. 

While this reconstruction formulation applies to general spatio-temporal datasets, our analysis focuses on stochastic realizations rather than a structured measurement grid.
As a result, individual realizations hold no physical significance on their own; instead, reconstruction quality is evaluated by comparing the average over all realizations, $X^{\rm Avg}$, to the average of the reconstructed trajectories, $X^{\rm Avg}_{\rm DMD}$.

\begin{figure}
\includegraphics[width=1\linewidth]{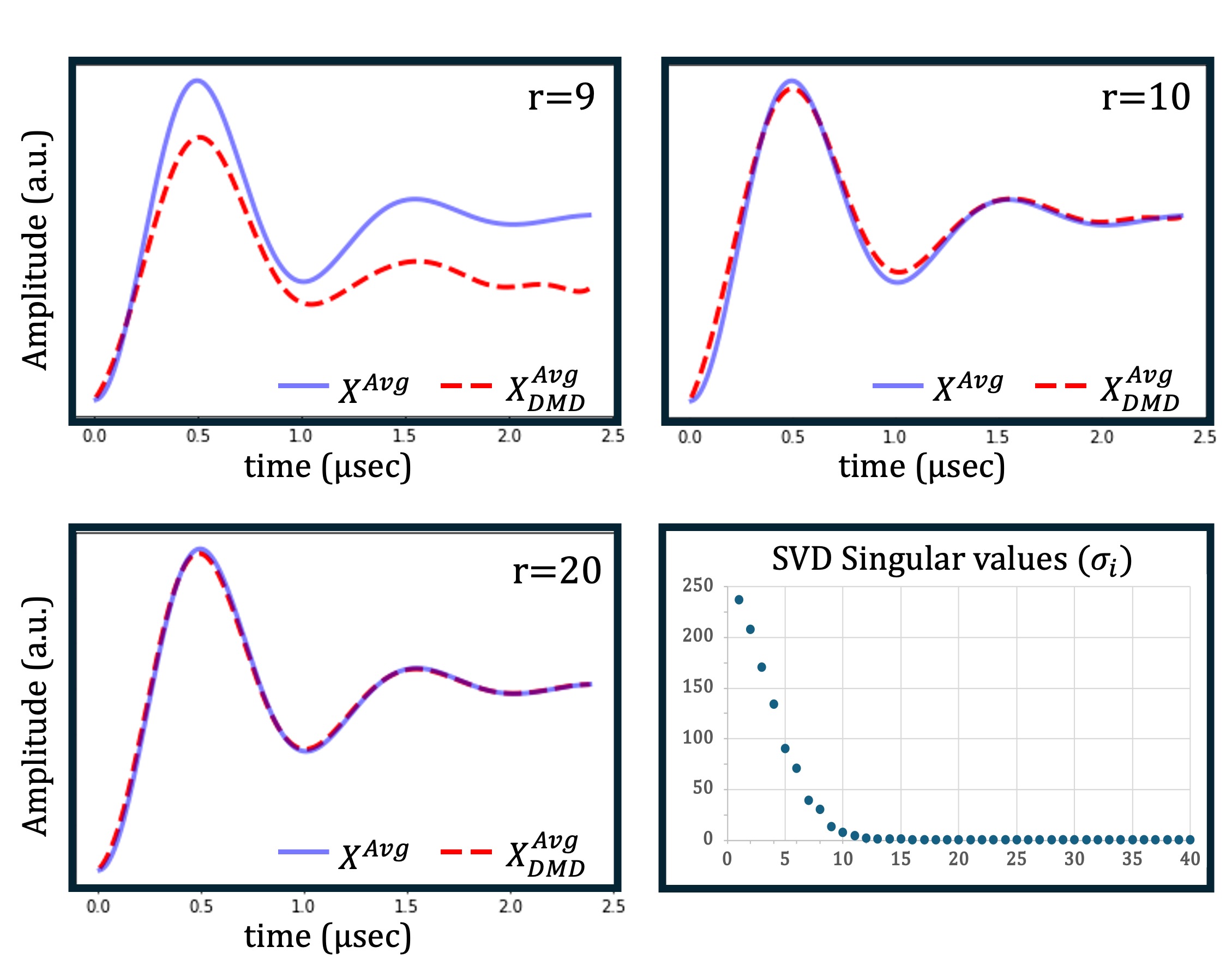} 
\caption{Comparison between the average over all stochastic realizations, $X^{\rm Avg}$ (solid blue line), and the average of the reconstructed DMD trajectories, $X^{\rm Avg}_{\rm DMD}$ (dashed red line), obtained from Eq.~(\ref{eq:DMD_reconstruct}), is shown for \textit{rank = 9} (upper-left), \textit{10} (upper-right), and \textit{20} (bottom-left). The case with \textit{rank = 20} yields the best alignment. The singular values from the SVD analysis reveal eight dominant components (bottom-right panel). Reconstruction with \textit{rank = 9} shows a large discrepancy between the reconstructed and input averages, while increasing the DMD rank to \textit{10} results in near-perfect alignment. This high sensitivity to rank selection reflects the stochastic nature of the input data.}
\label{fig:Fig-reconst}
\end{figure}

To ensure meaningful reconstruction under this averaging criterion, an appropriate rank $r$ for the DMD reduced space must be selected. The optimal rank is chosen to retain the minimum number of components necessary to capture the dominant physical phenomena. This is determined at the SVD level by selecting the $r$ largest singular values, $\{\sigma_i, \ldots, \sigma_r\}$, that capture the most significant features of the system's dynamics. To evaluate this choice, we compare $X^{\rm Avg}$ with $X^{\rm Avg}_{\rm DMD}$, as illustrated in Fig.~\ref{fig:Fig-reconst} for ranks $r=9, 10$, and $20$. Although the SVD singular values suggest that $r=9$ should be sufficient to capture the system's dynamics (bottom-right), the comparison between $X^{\rm Avg}$ (blue) and $X^{\rm Avg}_{\rm DMD}$ (dashed red) reveals a significant discrepancy, as shown in the top-left panel of Fig.~\ref{fig:Fig-reconst}. Increasing the rank to $r=10$ yields near-perfect alignment between the reconstructed and input averages, with perfect agreement achieved at $r=20$. This high sensitivity to rank selection is not reflected in the small difference between the singular values $\sigma_9$ and $ \sigma_{10}$, suggesting that the observed reconstruction discrepancies are driven by stochastic contributions that challenge standard DMD assumptions.

Achieving minimal reconstruction error with a reduced set of DMD components is essential before any physical interpretation can be made. In practical applications, the rank can be selected using data-internal diagnostics: the singular-value decay provides an initial estimate, while comparison between the measured ensemble average $X_{\rm Avg}$ and the ensemble-averaged DMD reconstruction $X^{\rm Avg}_{\rm DMD}$ identifies the smallest rank that reproduces the measured dynamics within the observation window. The robustness of the extracted spectral weights and decay time can then be tested by varying the rank around this value. The optimal rank depends on the measurement window, with longer durations generally requiring more components, an analysis of the rank-error dependence for different window durations is provided in App.~\ref{app:DMD}.

\section{Results}
\subsection{ Reinterpreting DMD for General Stochastic Dynamics}

Since DMD is designed for stationary systems, its direct application to our system, where fluctuations evolve dynamically, requires careful reconsideration. Standard DMD is commonly used to extract spatial-temporal modes from stationary systems with well-defined spatial-temporal structure, relying on the assumption that the underlying dynamics remain unchanged over time. However, our illustrative example is a two-level stochastic system that is neither stationary nor characterized by spatial modes. Thus, modifications to the DMD scheme are required to bridge this gap.
We first discuss the role of DMD's spatial modes within the stochastic framework, proposing a new interpretation. 

\begin{figure}
\includegraphics[width=1\linewidth]{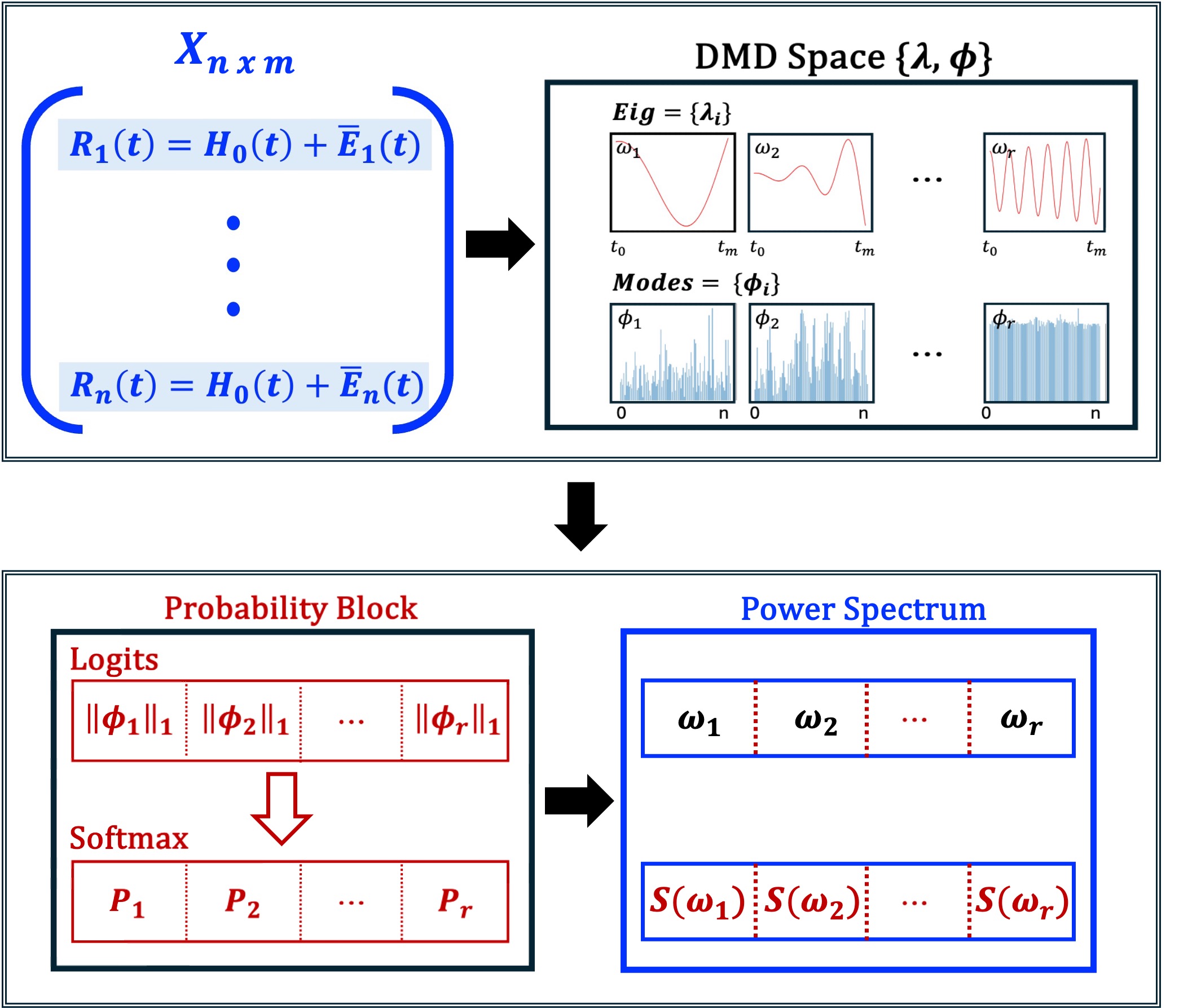} 
\caption{Schematic overview of our reinterpreted DMD-based noise characterization scheme. The upper block shows the input data matrix $X_{n \times m}$, composed of \textit{n} stochastic trajectories computed using Eq.~(\ref{eq:Hamiltonian}), each representing the expectation value of $\sigma_x$ over time. This matrix is decomposed into DMD eigenvalues and spatial modes $\{\lambda_i,\ \phi_i\}$, where each $\phi_i$ is interpreted as a vector in realization space. In the lower block, the $\ell_1$-norm of each mode,$\|\phi_i\|_1$, quantifies the total distribution of that mode across the ensemble. These raw values (serving as logits) are then passed through a softmax transformation to produce a peaked, normalized distribution $P_i$, 
which serves as a statistical weight for the associated
DMD-resolved angular frequency $\omega_i=\operatorname{Im}(\mu_i)$,
resulting in a data-driven spectrum $\{\omega_i,S(\omega_i)\}$.  
} 
\label{fig:Fig-noiseScheme}
\end{figure}

Figure \ref{fig:Fig-noiseScheme} displays the data flow in our scheme. The input matrix $X$ is derived from the stochastic simulation described above, where each row represents a different realization of the dynamics of the expectation value of $\sigma_x$ over the time interval $0\leq t \leq 7.0\ {\mu} s$. 
However, since no spatial coordinate is present, the DMD modes $\phi_i$ cannot be associated with a spatial distribution, requiring an alternative interpretation. Instead, each mode $\phi_i$ is an $n \times 1$ vector in realization space and can therefore be viewed as encoding how strongly the DMD component associated with $\omega_i$ is represented across the ensemble of realizations. 
The corresponding ensemble-level mode score can be quantified by its total magnitude, given by the complex $\ell_1$-norm of $\phi$:

\begin{equation}
    \|\phi\|_1 = \sum_{k=1}^{n} |\phi_k|= \sum_{k=1}^{n} \sqrt{\Re(\phi_k)^2+\Im(\phi_k)^2}.
    \label{1-norm}
\end{equation}

The raw $\ell_1\text{-norm}$ values ${\|\phi_i}\|_1$ quantify the contribution of each DMD mode \(\phi_i\) to the PSD-like amplitudes, but tend to be similar in magnitude due to the stochastic nature of the input. Each realization in the ensemble contains a different mixture of the system’s underlying coherent signal and random fluctuations. When DMD is applied to such data, it decomposes the noise across many modes with similar frequencies. These noise-driven modes are not specific to individual realizations but are shared statistically across the ensemble, forming a non-unique and overlapping basis. This behavior can be viewed as a form of effective degeneracy, analogous to degenerate modes in linear systems, where the dynamical contribution is spread across a non-unique set of modes with overlapping structure. As a result, the distribution of each mode across realizations appears broadly similar, leading to limited variation among their $\ell_1\text{-norm}$ and obscuring the underlying spectral structure.

To convert the raw mode contributions into spectral weights, we treat
$z_i=\|\phi_i\|_1$ as an unnormalized mode score for the contribution of the DMD frequency $\omega_i$ across the ensemble of stochastic realizations.
This is analogous to applying the softmax function in machine learning, where
unnormalized scores, or logits, are converted into positive normalized
weights. In the present setting, the desired output is not an absolute PSD,
but a normalized distribution of spectral weights over the DMD-resolved
frequencies. The softmax map provides such a distribution while preserving
the ordering of the raw mode scores and enhancing relative differences
between dominant and subdominant components.

Specifically, each value $z_i = \|\phi_i\|_1$ is mapped to a probability $p_i$ via the softmax function: 
\begin{equation} p_i = \frac{e^{z_i}}{\sum_{j} e^{z_j}} = \frac{e^{\|\phi_i\|_1}}{\sum_{j} e^{\|\mathbf{\phi_j}\|_1}} = S_i \label{eq-pi} \end{equation} 

The exponential form has two useful consequences. First, the weights are strictly positive and normalized, $\sum_i p_i=1$, so they can be interpreted as relative spectral weights. Second, ratios of weights depend exponentially on score differences,
\begin{equation}
\frac{p_i}{p_j}=\exp(z_i-z_j),
\end{equation}
so small differences between nearly degenerate DMD mode norms become visible without introducing a hard threshold. This property is useful for stochastic trajectory ensembles, where the raw mode norms can be close in magnitude because noise contributions are distributed over many modes. By substituting $z_i=\|\phi_i\|_1$, the DMD eigenpairs $\{\lambda_i,\phi_i\}$ are recast as a data-driven spectral-weight distribution $\{\omega_i,S(\omega_i)\}$.

We emphasize that this construction should be interpreted as a PSD-like spectral fingerprint rather than an absolute PSD estimator. The softmax normalization removes absolute power information and retains relative spectral weights over the finite set of DMD-resolved frequencies. Other nonlinear normalizations, including a temperature-scaled softmax $p_i(\beta)=e^{\beta z_i}/\sum_j e^{\beta z_j}$, could be used to tune the contrast of the spectral weights. Here we use the parameter-free choice $\beta=1$, which avoids introducing an additional fitting parameter and, as shown below, already separates the dominant spectral structure from the stochastic background.

\subsection{Noise Spectrum Extraction via DMD}

We applied the scheme illustrated in Fig.~\ref{fig:Fig-noiseScheme} to simulated data of a noisy quantum system with a characteristic frequency of $f_0=1$ \text{MHz}, validating our approach by computing the spectral weights of two different noise environments: $1/f$ noise and white noise. This comparison allows us to assess how well the method captures both broadband and low-frequency-dominated spectral features, and to evaluate its ability to extract physically meaningful signatures from finite, noisy datasets.
 
\begin{figure}
\includegraphics[width=1\linewidth]{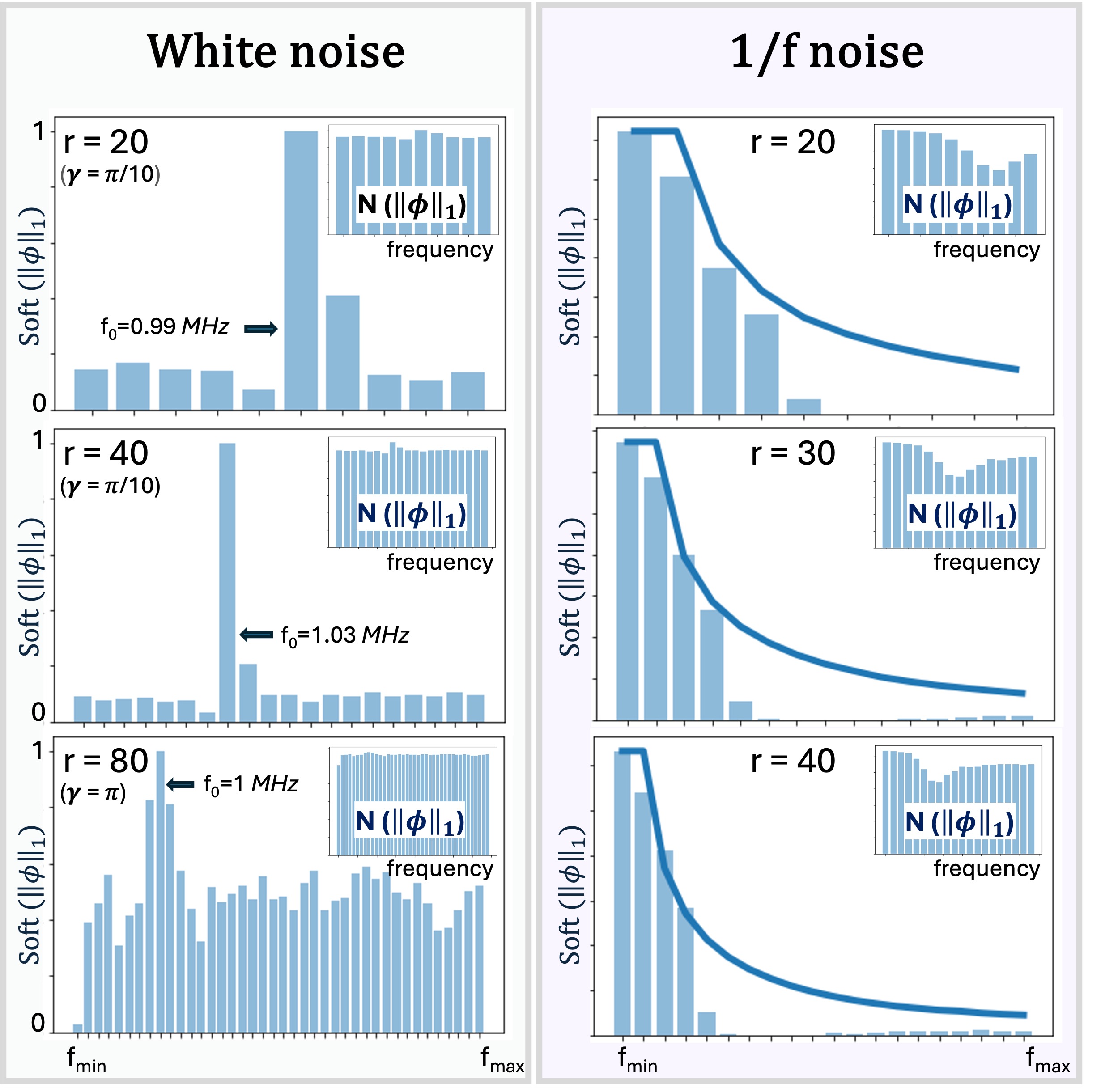} 
\caption{
The spectral weights extracted using our DMD-based scheme for a simulated quantum system with a characteristic frequency \(f_0 = 1\ \text{MHz}\), for white noise (left) and $1/f$ noise (right) environments. Spectral reconstructions were generated using DMD with selected ranks: 20, 40, and 80 for white noise, and 20, 30, and 40 for $1/f$ noise. The horizontal axis represents the natural frequencies
$f_i=\operatorname{Im}(\mu_i)/(2\pi)$ obtained from the
continuous-time DMD eigenvalues in Eq.~(3), and amplitudes correspond to softmax-transformed \(\ell_1\)-norms of the DMD modes. Insets show results using simple normalization for comparison. For white noise, softmax reveals a sharp spectral peak at \(f_0\) under weak noise \((\gamma = \frac{\pi}{10}~\text{rad}/\mu s)\), which becomes masked under stronger noise \((\gamma = \pi~\text{rad}/\mu s)\) or with simple normalization. For \(1/f\) noise, the spectral weights reveal dominant low-frequency structure aligning with the overlaid inverse-frequency trend (solid blue line), which is not recovered with simple normalization.
}
\label{fig:Fig-Softmax-white-pink-noiseScheme}
\end{figure}

Results are shown in Fig.~\ref{fig:Fig-Softmax-white-pink-noiseScheme} for reduced DMD space at ranks 20, 40, and 80 for white noise (left), and ranks 20, 30, and 40 for $1/f$ noise (right), respectively. The horizontal axis in each panel represents the natural frequencies
$f_i=\operatorname{Im}(\mu_i)/(2\pi)$, where the continuous-time
DMD eigenvalues $\mu_i$ are obtained from the discrete eigenvalues
$\lambda_i$ using Eq.~(3). As the rank increases, $f_{max}$ also increases due to the inclusion of singular vectors corresponding to smaller singular values, which capture finer details and smaller-scale features in the time-series data~\cite{baratz2024unsupervised}. For each frequency $f_i$, the corresponding amplitude is given by the normalized weight $S_i$ in Eq.~(\ref{eq-pi}).

We begin by comparing the power spectrum of white noise obtained from our scheme with a simple normalization of the $\ell_1\text{-norm}$ values, $\|\phi_i\|_1$, defined as $N(\|\phi_i\|_1)=\frac{\|\phi_i\|_1}{\sum_{j} \|\mathbf{\phi_j}\|_1}$ as shown in the inset of each panel. This comparison is presented in the left panel of Fig.~\ref{fig:Fig-Softmax-white-pink-noiseScheme} for two white noise amplitudes: weak ($\gamma=\frac{\pi}{10}~\text{rad}/\mu s$) and strong ($\gamma=\pi~\text{rad}/\mu s$). 

In the weak noise case, the amplitudes of all DMD frequencies are nearly uniform, except for a single frequency aligned with the system's characteristic frequency $f_0$, which stands out with significantly higher amplitude. This is the expected spectral signature of a system dominated by a single coherent mode, while the white noise introduces a uniform, lower amplitude background across the remaining frequencies. Notably, this pattern is clearly revealed when applying the softmax transformation to the set $\{{\|\phi_i\|_1}\}$, but it is almost entirely lost under simple linear normalization. 

In the case of strong white noise (left panel, bottom), the system frequency, now masked by stronger fluctuations, is divided among several DMD modes. This is expected, as strong noise obscures the system's coherent dynamics, forcing DMD to represent the signal using a broader set of modes. In this case, the spectral structure completely vanishes under simple normalization, whereas the softmax transformation still reveals its peak-like feature.

For $1/f$ noise, the spectral-weight distribution reveals dominant low-frequency structure aligning with the overlaid inverse-frequency trend (solid blue line), which is not recovered with simple normalization. To further assess its resemblance to a true $1/f$ noise spectrum, we apply Eq.~(\ref{eq-omega}) to the list of DMD eigenvalues $\{\lambda_1,..,\lambda_{rank}\}$ to compute the corresponding frequencies $\{f_1,...,f_{rank}\}$, followed by taking their reciprocals. This trend is shown as the solid blue line in each panel. As seen, the histogram amplitudes for the low-lying frequencies align with the $1/f$ trend depicted by the blue line. This pattern is not observed using a simple normalization, as shown in the insets of the right panel. Since the magnitude of the noise in this case is equivalent to the strong white noise scenario (i.e., $\gamma=\pi~\text{rad}/\mu s$), the system frequency is no longer visible, as it is completely masked by the high amplitudes of low-frequency noise components with their strong long-term correlations. Notably, a $1/f$ spectral behavior typically emerges when a broad range of correlated frequency components contribute to the system's dynamics, a condition that is often difficult to verify experimentally due to data limitations. This makes our DMD scheme a valuable tool for extracting the system’s actual noise characteristics from limited data.

\subsection{Comparison with True $1/f$ Noise}

To further validate our method, we compared the DMD-derived $1/f$  spectral weights with the true $1/f$ PSD used in the simulations. However, a direct comparison is not straightforward, as the DMD eigenvalues are constrained by the chosen rank and represent generalized dynamics in a reduced space. For instance, with rank=$20$, DMD yields twenty eigenvalues, corresponding to only ten distinct natural frequencies (since each natural frequency is associated with a conjugate eigenvalue pair). 
In contrast, the true $1/f$ noise spectrum is computed directly using 500 fluctuators (see App.~\ref{app:1_over_f}). 
To bridge this gap, we introduce a comparative approach that links the low-rank DMD $1/f$ spectrum with the actual $1/f$ noise spectrum underlying the simulated dataset by drawing an analogy between DMD rank selection and a filtering operation. The rank selection in DMD is guided by the magnitude of the singular values, which represent the energy of the singular vectors in descending order. 

By appropriately choosing the rank, we effectively filter out high-frequency components from the DMD-reduced space. This operation can be viewed as applying a low-pass filter to the measured $1/f$ noise. To achieve this, we convolved the actual $1/f$ spectrum taken from the simulation with a Gaussian function. The full width at half maximum (FWHM) of this Gaussian was chosen to reflect the ratio between the simulation bandwidth, i.e., the full frequency range used to generate the $1/f$ noise, and the number of DMD natural frequencies, which is dictated by the DMD rank. This relationship is defined as follows:
\begin{equation}
    \text{FWHM} = \frac{\Delta f_{in}}{(0.5\cdot rank)},
\label{eq-FWHM}    
\end{equation}
where $(0.5\cdot rank)$ accounts for the number of distinct DMD natural frequencies while $\Delta f_{in}$ denotes the total frequency bandwidth defined in the simulation. This smoothing process adjusts the true $1/f$ spectrum to align with the spectral resolution determined by the DMD rank, enabling a direct one-to-one comparison between the two spectra.

\begin{figure}
\includegraphics[width=1\linewidth]{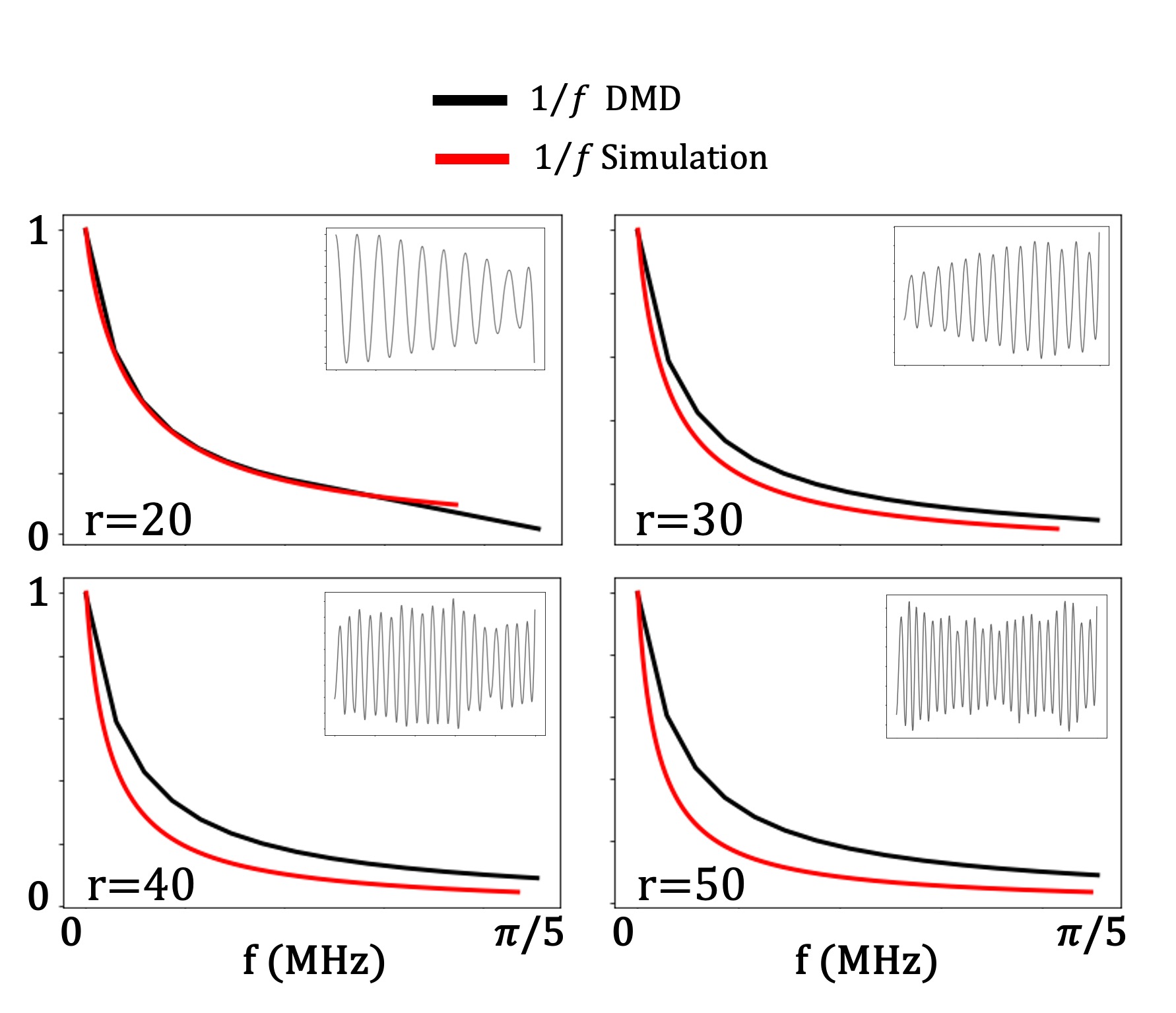} 
\caption{Comparison between the DMD-derived \(1/f\) spectrum (black) and a Gaussian convolution of the true (simulated) \(1/f\) noise spectrum (red).
 The DMD-derived spectral weights are compared with an inverse-frequency trend evaluated at the DMD-derived natural frequencies, while the true simulated spectrum is convolved with a Gaussian whose full width at half maximum (FWHM) is given by Eq.~(\ref{eq-FWHM}).}
This convolution mimics a filtering effect that bridges the gap between the high resolution of the true spectrum and the lower spectral resolution imposed by the DMD rank. Results are shown for DMD ranks 20, 30, 40, and 50. The closest agreement is observed at rank 20, consistent with SVD analysis indicating ten dominant singular values, which produce 20 DMD modes. At higher ranks, discrepancies increase due to the inclusion of vectors associated with near-zero singular values, which introduce high-frequency components. Insets highlight the highest frequency captured in each reduced space, illustrating the effective low-pass filtering imposed by DMD rank selection.
\label{fig:Fig-convolution}
\end{figure}

Fig.~\ref{fig:Fig-convolution}  compares the DMD-derived spectral weights with an inverse-frequency trend evaluated at the DMD-derived natural frequencies, together with the convolution of the true $1/f$ noise using a Gaussian function with a FWHM determined by Eq.~(\ref{eq-FWHM}). The comparison is shown for ranks 20, 30, 40, and 50. The best agreement is observed at $rank=20$, suggesting that a reduced space with ten natural frequencies is optimal for capturing the system's dynamics. This finding aligns with the SVD analysis, where the first ten singular values are substantially larger than the remaining ones, as discussed above and shown in the bottom-right panel of Fig.~\ref{fig:Fig-reconst}. As the rank increases, the deviation between the two spectra grows due to the inclusion of singular vectors with near-zero singular values, which introduce high-frequency components into the DMD-reduced space. To further highlight the low-pass filtering effect, the highest frequency component included in each reduced space is depicted in the inset of each panel. \\ 

To place our DMD-based noise reconstruction in context, we benchmarked it
against several conventional noise-spectroscopy approaches using the same
Ramsey-type trajectories (see Supplementary Material).
Fourier-based PSD estimators (Welch and related methods) applied directly
to $\langle\sigma_x(t)\rangle$ give a robust estimate of the qubit
oscillation frequency and very coarse information about whether the noise
is broadband or low-frequency dominated, but they do not directly yield
the bath spectral weights $S(\omega)$. Forward-model inversions of the
filter-function relation, including Tikhonov and more elaborate Bayesian
schemes, can in principle reconstruct $S(\omega)$ and provide
uncertainty estimates or band-limited noise strengths, but only when an accurate analytic dephasing model is available, suitable priors are chosen, and often when tailored control sequences are used. Harmonic inversion and filter-diagonalization methods are highly effective for resolving discrete spectral lines, yet they naturally return effective poles of the dephasing envelope rather than a broadband bath spectrum.
In contrast, the proposed DMD framework operates directly on stochastic trajectories, requires no parametric model for $S(\omega)$ or specialized control, and produces a PSD-like spectrum that more faithfully reproduces the simulated white and $1/f$ noise over the accessible bandwidth using the same short time windows and a modest number of realizations. 
A more detailed benchmark against Welch spectral estimators and Tikhonov-regularized forward-model inversion is provided in the Supplementary Material, together with a discussion of the relation to harmonic inversion, filter-diagonalization, and Bayesian noise-spectroscopy approaches.

Moreover, as we show next, the same DMD analysis also yields the intrinsic coherence time and enables constrained extrapolation of the dynamics beyond the measurement window.

\subsection{Decoherence Time Extraction}

Next, we demonstrate how collective decay or coherence times can be extracted from the DMD eigenvalues, illustrated explicitly through the quantum-specific decoherence time $T_2^*$. Similar concepts apply to classical systems exhibiting characteristic decay dynamics. 
The set of complex DMD eigenvalues characterizes both the oscillatory behavior
and the growth or decay dynamics of the system: the imaginary part of
$\mu_i$ gives the angular frequency, $\omega_i=\operatorname{Im}(\mu_i)$, while the real part gives the growth or decay rate.
Given that pure dephasing ($T_2^*$) is the only decay process in our simulations, this coherence decay can be identified with the DMD mode associated with a real eigenvalue $\lambda_i$.
For the DMD mode associated with the dephasing envelope, we define
$
T_2^*=-\frac{1}{\operatorname{Re}(\mu_{T_2^*})}
=
-\frac{\Delta t}{\log|\lambda_{T_2^*}|}.
$

To enable the extraction of such a real eigenvalue, we perform DMD analysis with an odd rank. Since each oscillatory mode corresponds to a conjugate pair of eigenvalues, choosing an odd rank allows DMD to isolate a single real eigenvalue that captures the ensemble's phase decay as a distinct non-oscillatory mode. This is demonstrated by comparing the dynamics of the mode associated with the real eigenvalue to the system's stochastic evolution, represented by the average over all realizations. 

As shown in the top panel of Fig.~\ref{fig:Fig-FuturePr}, the decay captured by this mode (gray dashed line) closely aligns with the amplitude decay observed in the system's evolution (black solid line). This suggests that the real eigenvalue obtained through DMD provides a reliable approximation of the coherence decay governed by $T_2^*$.
Notably, $T_2^*$ cannot be directly extracted from the raw stochastic trajectories without assuming a specific functional form, such as exponential or Gaussian decay, to fit the coherence envelope. In contrast, its emergence as a distinct dynamical component in the DMD eigenvalue spectrum is a key outcome uniquely enabled by our data-driven approach.

\subsection{Stabilized Dynamics Extrapolation}

\begin{figure}
\includegraphics[width=1.0\linewidth]{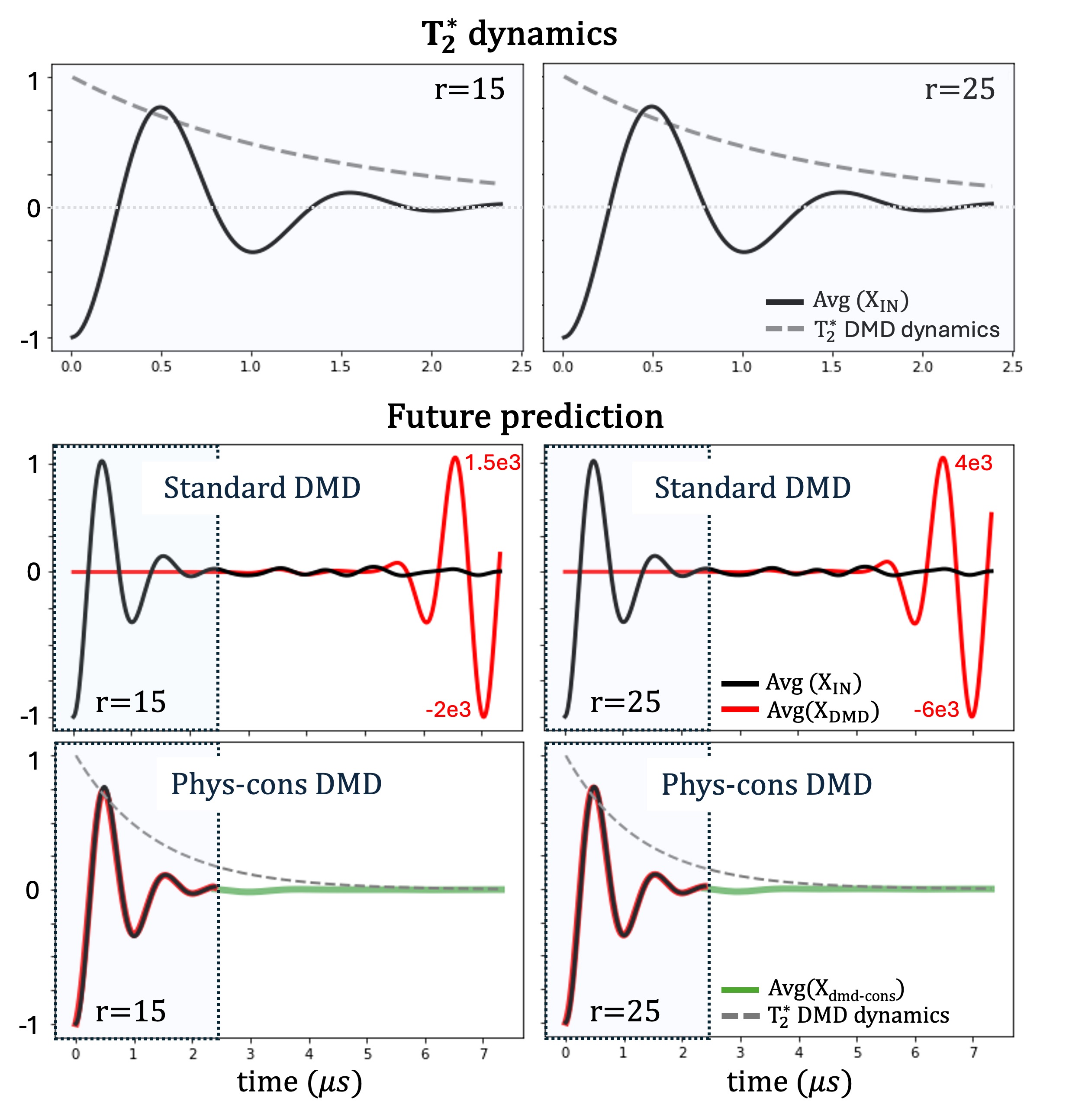} 
\caption{\textbf{Top panel:} Average stochastic evolution of the system (black solid line) compared with the dynamics of the DMD mode associated with \(T_2^*\), given by \(b_{T_2^*} e^{\mu_{T_2^*} t}\) (gray dashed line). Results are shown for two odd DMD ranks: \(R = 15\) (left) and \(R = 25\) (right). \textbf{Middle panel:} True average evolution of the system \(X^{\text{Avg}}(t)\) over the full interval \(0 \leq t \leq 7.0\ \mu s\) (black line), compared with standard DMD extrapolation (red) using Eq.~(\ref{eq:DMD_reconstruct}). The DMD analysis was applied only on the analysis window \(0 \leq t \leq 2.5\ \mu s\), indicated by the light gray shaded region. 
While the standard DMD extrapolation initially tracks the true dynamics within the analysis window, it diverges significantly beyond $t>5~\mu s$. At long times, the red curve reaches amplitudes approximately $10^3$ larger than the physical signal, as indicated in the panel, due to unstable high-frequency modes with eigenvalues slightly above one.
\textbf{Bottom panel:} Constrained DMD prediction (green) computed using Eq.~(\ref{eq:futurePred})  for $t>2.5\ \mu s$, shown together with the ground-truth average dynamics (black) reconstructed within the analysis window. Since the constrained prediction closely overlaps the reference dynamics outside the reconstruction window, the black curve is largely covered by the green curve over much of the plotted interval. The corresponding reference dynamics are therefore also shown explicitly in the middle panel in black. The constrained reconstruction imposes a physical bound on eigenvalue magnitudes based on the $T_2^*$ decay rate and replaces initial-condition weights $b_i$ with spectral weights $S_i$. The dotted gray line shows the exponential decay associated with the $T_2^*$ DMD mode, obtained from standard DMD extrapolation (Eq.~(\ref{eq:DMD_reconstruct})). The constrained prediction remains stable and closely follows the true dynamics beyond the measurement window.}
\label{fig:Fig-FuturePr}
\end{figure}

One of the key advantages of DMD is its ability to predict dynamical behavior beyond the temporal analysis window of the input data. This results from the exponential time dependence of the DMD modes, see Eq.~(\ref{eq:DMD_reconstruct}), which allows the analysis to extrapolate future states of the system under the assumption that the underlying dynamics remain unchanged~\cite{Marusic_2024}. However, in the presence of noise, this assumption no longer holds. Stochastic fluctuations disrupt the eigendecomposition of the system over time, preventing reliable extrapolation beyond the measurement window. This effect is demonstrated in the middle panels of Fig.~\ref{fig:Fig-FuturePr}. The system evolution given by $X^\text{\rm Avg}$ across the full time range $0 \leq t \leq 7\ \mu s$ is shown in black for \textit{R}=15 (left) and \textit{R}=25 (right).

In contrast, the DMD analysis was performed using only a shorter temporal window of $0 \leq t \leq 2.5\ \mu s$, indicated by the light gray shaded region. From this limited dataset, we applied Eq.~(\ref{eq:DMD_reconstruct}) to extrapolate the system’s dynamics, extending up to $t=7\ \mu s $ shown in red. To enable direct comparison, the amplitudes of both the true (black) and predicted (red) dynamics were normalized to one. 
While the standard DMD extrapolation initially tracks the true dynamics within the analysis window, it begins to diverge significantly beyond $t>5~\mu s$. It ultimately displays instability, with amplitudes growing by approximately three orders of magnitude relative to the physical signal, as annotated in red near the peaks.

The source of this instability lies in certain high-frequency DMD modes with eigenvalues slightly greater than one, causing their contributions to grow over time rather than decay. Within the analyzed temporal window, the overall contribution of these high-frequency components remains small due to the $1/f$ PSD nature of the noise, which emphasizes low-frequency components and thus results in an overall decaying trend. However, when extrapolating beyond this window, the well-behaved low-frequency components, those with eigenvalues of magnitude less than one, have already decayed. As a result, they can no longer counterbalance the growth of spurious high-frequency components. Due to the exponential nature of the resolved dynamics, these minor contributions from unstable modes accumulate over time, ultimately leading to instability in the predicted system behavior.\\

To overcome this instability, we impose a physically motivated constraint on the eigenvalue magnitudes entering the extrapolated dynamics. The standard DMD reconstruction evolves each mode as $e^{\mu_i t}$, so the real part of $\mu_i$, or equivalently the magnitude of the corresponding discrete-time eigenvalue $\lambda_i$, determines whether that mode decays or grows during extrapolation. In the noisy finite-window data considered here, some weak high-frequency modes acquire eigenvalue magnitudes that are inconsistent with the observed ensemble relaxation. These modes may have little influence inside the analysis window, but can dominate the prediction at later times.

In the dephasing problem, the ensemble-averaged coherence is expected to remain bounded and decay according to the coherence envelope. The DMD spectrum identifies this envelope through the real decay component associated with $T_2^*$. 

We therefore use the magnitude of the eigenvalue associated with this component, denoted  $|\lambda_{T_2^*}|=\exp(-\Delta t/T_2^*)$, as a data-driven physical bound on the eigenvalue magnitudes. This constraint is stronger than simply requiring stability: it suppresses modes that would either grow outside the measurement window or decay more slowly than the extracted $T_2^*$ envelope. Any eigenvalue exceeding this threshold is rescaled accordingly, while preserving its phase, and we define the adjusted eigenvalue as $\lambda^*_i=\left(\frac{\lambda_i}{|\lambda_i|}\right) |\lambda_{T_2^*}|$.

For the predictive ensemble reconstruction, we also replace the standard DMD amplitudes $b_i$ by the spectral weights $S_i$ learned from the stochastic ensemble, as described by Eq.~(\ref{eq-pi}). The coefficients $b_i$ are chosen in standard DMD to reproduce the initial condition of the training data. Here, however, the goal is to predict the ensemble-averaged long-time behavior using the spectral content inferred within the analysis window. The weights $S_i$ therefore provide an ensemble-level weighting of the DMD-resolved frequencies, consistent with the PSD-like spectral representation introduced above.

Both modifications are incorporated into a constrained reconstruction scheme, Eq.~(\ref{eq:futurePred}), which describes the predicted dynamical behavior under these constraints. Here, $X_{dmd-cons}(t)$ represents the system’s state at future times $t$ beyond the analysis window, and $\mu^*_i$ is the constrained continuous-time DMD eigenvalue derived from the adjusted discrete-time DMD eigenvalue $\lambda^*_i$.

\begin{equation}\label{eq:futurePred}
    X_{dmd-cons}(t) \approx \sum_{i=1}^{r} S_i \cdot e^{\mu^*_i t}
\end{equation}
\begin{equation*}
    \mu^*_i =
    \begin{cases} 
        \frac{\log(\lambda^*_i)}{\Delta t}, & \text{if } |\lambda_i| > |\lambda(T^*_2)|  \\[8pt]
        \mu_i, & \text{if } |\lambda_i| \leq |\lambda(T^*_2)|
    \end{cases}.
\end{equation*}

Using this formulation, we apply Eq.~(\ref{eq:futurePred}) to predict the system's dynamics beyond the measurement window, in the range $2.5 \leq t \leq 7.0 \ \mu s$, as shown by the green line in the bottom panel of Fig.~\ref{fig:Fig-FuturePr} for $R=15$ (left) and $R=25$ (right). The light gray shaded region indicates the measurement window, where the standard DMD reconstruction (red), computed with Eq.~(\ref{eq:DMD_reconstruct}), is shown alongside the average evolution across all trajectories (black).
The gray dotted line is the decay predicted by the $T^*_2$ DMD component over the full interval $0 \leq t \leq 7.0\ \mu s$, demonstrating that the constraint imposed by Eq.~(\ref{eq:futurePred}) effectively suppresses the long-time instability of the standard DMD extrapolation, as reflected in the asymptotic alignment of the green and gray dotted lines. Furthermore, incorporating $S_i$ as the new weighting factors in the linear combination of DMD modes ensures that the predicted dynamics not only remain stable, but also transition smoothly from the behavior captured within the measurement window. 
Since each $S_i$ represents an ensemble-level spectral weight learned within the analysis window, the extrapolated dynamics are weighted by the stochastic spectral content of the data rather than by the initial-condition amplitudes alone.
Notably, the green constrained prediction overlaps closely with the ground-truth average dynamics in the bottom panel. Since this overlap can obscure the black reference curve, the same ground-truth dynamics are also shown explicitly in the middle panel, allowing direct comparison with both the standard and constrained DMD predictions.

\section{Discussion}

We introduced a general, fully data-driven framework for characterizing stochastic dynamics and noise directly from ensembles of short, noisy trajectories using DMD. By reinterpreting DMD’s spatiotemporal modes as ensemble-level statistical weights and mapping them through a nonlinear softmax transformation, we obtain a normalized spectral-weight distribution, a PSD-like spectral fingerprint, that captures the dominant frequency content of the underlying stochastic fluctuations.

This spectral fingerprint should be interpreted as a finite-data descriptor: its resolution and robustness depend on the available trajectory ensemble, the sampling interval, the measurement window, the noise level, and the selected DMD rank. 
For data with strongly drifting or explicitly time-dependent statistics, the same DMD construction may need to be applied in moving windows or adaptive variants, which we leave for future work.
In experimental applications, measurement noise enters as part of the finite-data trajectory ensemble, and the robustness of the extracted features can be assessed by checking reconstruction accuracy within the measurement window and stability under moderate changes of rank or analysis window.
With these finite-data qualifications, the DMD-derived spectral-weight representation remains physically interpretable and discriminates between broadband (white) and correlated ($1/f$) environments without assuming a specific parametric form for the noise spectrum.

Beyond spectral characterization, the same decomposition yields intrinsic dynamical time scales: the DMD eigenvalues encode coherent oscillations and decay rates, enabling direct extraction of characteristic coherence and relaxation times from time-domain data. In our qubit dephasing case study, this allows coherence-time estimation and noise profiling from limited records, in regimes where conventional spectral estimators require substantially longer and cleaner time records and inverse-problem approaches depend sensitively on priors and regularization. Finally, to overcome the well-known instability of naive DMD extrapolation in noisy, nonstationary settings, we introduced a constrained predictive reconstruction that anchors long-time behavior using physically motivated decay bounds together with the learned spectral weights. This stabilizes forward prediction while retaining the model-free nature of the approach.

Our results establish DMD as a unified data-driven framework for (i) extracting spectral structure of the noise, (ii) identifying intrinsic decay and coherence times, and (iii) enabling stable prediction beyond the observation window, using only time-series data and minimal assumptions. 
While the present validation should be regarded as a controlled synthetic-data benchmark performed on simulated qubit dephasing dynamics, the DMD construction itself is formulated at the level of stochastic trajectory ensembles.
Once a measured observable is represented as a collection of time traces, the analysis does not depend on whether the underlying stochastic dynamics originate from a quantum or classical system. A simple classical analogue of the model studied here is a phase oscillator with stochastic instantaneous frequency: writing $u(t)=e^{i\theta(t)}$ and $\dot{\theta}(t)=\omega_0-\xi(t)$ gives $\dot{u}(t)=i[\omega_0-\xi(t)]u(t)$, which has the same stochastic phase-accumulation structure as the transverse coherence in the qubit dephasing model. Thus, the qubit model serves as a controlled benchmark, while related stochastic phase-noise problems in classical oscillators, electronic circuits, optical systems, or mechanical resonators provide natural extensions of the same trajectory-level framework~\cite{Kubo1969,Demir2000}.

\section*{Acknowledgments}
AA.B and S.R-A acknowledge support by the Weizmann Artificial Intelligence Institute and Hub. This research was supported by the ISF Grants No.~1364/21, 3105/23, 700/22, and 702/22. A.H. has been generously supported by Dr. Arik Carasso, Honorary Doctor. L.M.C acknowledges the CINECA award under the	 ISCRA initiative, for the availability of high-performance computing resources and support (IscraC ``IsCc6\_CTHNEQS", IscraB	 ``IsB31\_EQDNPH").

\printcredits

\appendix

\section*{Appendix}
\section{Numerical simulations of qubit dynamics with noise }

\subsection{1/f noise}
\label{app:1_over_f}
Over the years, the study of $1/f$ noise in condensed matter systems and quantum information devices \cite{bergli2009decoherence}, has been tackled by means of a range of theoretical methods \cite{paladino20141}. Below, we choose to model the decoherence effects arising from $1/f$ noise by making use of the spin-fluctuator model \cite{paladino2002decoherence,falci2005initial,bergli2009decoherence,burkard2009non}. Within this theoretical framework, the properties of the noise are modeled by means of a collection of $N_{\text{F}}$ two-level classical background fluctuators. Each one of them undergoes a random telegraph process $\xi_{i}(t)$ \cite{Gardinerbook}, where we denote the symmetric switching rates with $\gamma_{i},i=1,\dots,N_{\text{F}}$. This is also known as Random-Telegraph noise \cite{haunggi1994colored}. A minimal model Hamiltonian describing the dynamics of a single quantum degree of freedom under the influence of such a kind of noise can be written as follows 
\begin{equation}\label{eq:1/f}
H(t)=H_{\text{o}} - \frac{1}{2}\xi(t)\sigma_{z},
\end{equation}
where $H_{\text{o}}$ describes a quantum two-level system, i.e., a qubit, modeled by means of Pauli operators $\sigma_{\alpha}\mbox{, }\alpha=x,y,z$. Hamiltonians of the form of Eq.~(\ref{eq:1/f}) have been known for a long time as generalizations of the Kubo-Anderson model \cite{Blume68}.

Below, we focus on the simulation of the qubit dynamics in the limit of pure dephasing, i.e., $H_{\text{o}}=\omega_{o}\sigma_z/2$. Moreover, we set $\xi(t)=\sum_{i=1}^{N_{\text{F}}}\xi_{i}(t)$, where each stochastic variable $\xi_{i}(t)$ switches between two discrete values $\pm v$. They can be equivalently expressed as follows \cite{haunggi1994colored}
\begin{equation}\label{eq:1/f2}
\xi_{i}(t)=v_{i}(-1)^{N_{i}(t)}, 
\end{equation}
where $N_{i}(t)$ is a further stochastic variable that obeys Poisson statistics with average $\ev{N_{i}(t)}=\gamma_{i}t$. It follows that, for each $\gamma_{i}$ the average exponentially decays to its stationary value $\ev{\xi_{i}(t)}=v_{i} e^{-2\gamma_{i}t}$. In what follows, we are interested in the stationary autocorrelation function, which reads
\begin{equation}\label{eq:1/f3}
\ev{\xi_{i}(t)\xi_{i}(s)}_{\text{st}}=v^2_{i} e^{-2\gamma_{i}\abs{t-s}}.
\end{equation}
We also assume the different variables to be uncorrelated, so that $\ev{\xi(t)\xi(s)}_{\text{st}}\simeq \sum_{i}v^2_{i}e^{-2\gamma_{i}\abs{t-s}}$.
For a single fluctuator of fixed switching rate $\gamma$, the noise spectrum \cite{GardinerZoller} has a Lorentzian shape 
\begin{multline}\label{eq:1/f4}
S_{\text{sing}}(\omega)= \int_{-\infty}^{+\infty}\mathrm{d}\tau e^{-i\omega \tau}\ev{\xi(\tau)\xi(0)}_{\text{st}}=\\ =v^2\int_{-\infty}^{+\infty}\mathrm{d}\tau e^{-2\gamma\abs{\tau}-i\omega \tau}=v^2 \frac{4\gamma}{4\gamma^2 +\omega^2}.
\end{multline}
To simulate $1/f$ noise, we now consider a set of $N_{\text{F}}$ fluctuators, where the corresponding $\gamma_i$ are drawn from a probability distribution of the form $P(\gamma)\propto 1/\gamma $ \cite{falci2005initial,bergli2009decoherence,burkard2009non}. For the sake of simplicity, we assume each fluctuator to be equally coupled to the controlled two-level system, i.e., $v_{i}=\bar{v}$. The power spectrum of the noise can be computed as
\begin{multline}\label{eq:1/f5}
S^{\text{1/f}}(\omega)=\int_{\gamma_{\text{o}}}^{\gamma_{\text{c}}}\mathrm{d}\gamma P(\gamma)\bar{v}^2 \frac{4\gamma}{4\gamma^2 +\omega^2}=\\=\frac{4 \bar{v}^2}{\log(\gamma_{\text{c}}/\gamma_{\text{o}})}\int_{\gamma_{\text{o}}}^{\gamma_{\text{c}}}\mathrm{d}\gamma  \frac{1}{4\gamma^2 +\omega^2}=\\=\frac{2 \bar{v}^2}{\log(\gamma_{\text{c}}/\gamma_{\text{o}})}\frac{1}{\omega}\qty(\arccot(\frac{2\gamma_{\text{o}}}{\omega})-\arccot(\frac{2\gamma_{\text{c}}}{\omega})),  
\end{multline}
where we denoted with $[\gamma_{\text{o}},\gamma_{\text{c}}]$ the cutoff frequencies. In the limit $\gamma_{\text{o}}\to 0^{+}$, we get that the low-frequency behavior of the spectrum is $\propto 1/\omega$.
The spectrum in Eq.~(\ref{eq:1/f5}) can be mimicked by a finite set of fluctuators $N_{\text{F}}$ as follows
\begin{equation}\label{eq:1/f6}
S^{\text{1/f}}(\omega)\simeq\sum_{i=1}^{N_{\text{F}}}v^2_{i} \frac{4\gamma_{i}}{4\gamma^2_{i} + \omega^2}=N_{\text{F}}\bar{v}^2 \langle\langle\frac{4\gamma}{4\gamma^2 +\omega^2}\rangle\rangle 
\end{equation}
where we denoted with $[\gamma_{\text{o}},\gamma_{\text{c}}]$ the cutoff frequencies. In the limit $\gamma_{\text{o}}\to 0^{+}$, we get that the low-frequency behavior of the spectrum is $\propto 1/\omega$. 
The spectrum in Eq.~(\ref{eq:1/f5}) can be mimicked by a finite set of fluctuators $N_{\text{F}}$ as follows
\begin{equation}\label{eq:1/f6}
S^{\text{1/f}}(\omega)\simeq\sum_{i=1}^{N_{\text{F}}}v^2_{i} \frac{4\gamma_{i}}{4\gamma^2_{i} + \omega^2}=N_{\text{F}}\bar{v}^2 \langle\langle\frac{4\gamma}{4\gamma^2 +\omega^2}\rangle\rangle 
\end{equation}
where the double brackets denote the average over $P(\gamma)$. A rather easy way to simulate Eq.~(\ref{eq:1/f5}) is thus to take a sufficient number of fluctuators distributed according to $P(\gamma)$, keeping the factor $N_{\text{F}}\bar{v}^2$ finite for each different spectrum width $\gamma_{\text{c}}-\gamma_{\text{o}}$. Taking advantage of the form in Eq.~(\ref{eq:1/f2}), a single realization of $\xi(t)$ is simulated using $N_{\text{F}}$ distinct Poisson variables, which experience jumps at random waiting times $t_{k}$ in the course of the time $[0,t_f]$. Each jump results in a sudden fluctuation of the Hamiltonian in Eq.~(\ref{eq:1/f}). The dynamics of the pure state $\ket{\psi_{\xi}(t)}$ corresponding to each distinct realization $\xi$ can be simulated by means of deterministic algorithms in between consecutive random jumps. After each jump, the Hamiltonian in Eq.~(\ref{eq:1/f}) is updated, and the dynamics is then restarted from the state right before the jump. The initial state of the qubit is set to a superposition state $\ket{\psi(0)}=1/\sqrt{2}(\ket{0}-\ket{1})$, which is the same for each realization. On the other hand, the initial condition of the noise variable, i.e., $\xi(0)$, can be different from one realization to the other \cite{paladino20141}. We thus extract the init state of each variable $\xi_{i}(t)$ from initial distribution with probability $P(\xi_{i}(0)=\pm v)=1/2$, $i=1,\dots N_{\text{F}}$.

The expectation value of qubit observables can thus be computed for each distinct realization of $\xi(t)$, i.e., $\ev{O(t)}_{\xi}=\bra{\psi_{\xi}(t)}O\ket{\psi_{\xi}(t)}$. Moreover, the density matrix can be computed at any time $t$, i.e.,
\begin{equation}\label{eq:density}
\rho(t)=\overline{\ketbra{\psi_{\xi}(t)}},
\end{equation}
where we denoted with $\overline{O}$ the average over a sufficient number of realizations of the noise $\xi(t)$.

\subsection{White noise}
\label{app:white_noise}
White noise spectrum arises from delta-correlated stochastic variables \cite{Gardinerbook} of the form
\begin{equation}\label{eq:white1}
\ev{\xi(t)\xi(s)}_{\text{st}} =\gamma \delta(t-s),   
\end{equation}
where $\gamma$ plays the role of the noise strength. To simulate the resulting stochastic Schr\"odinger equation \cite{gordon2008optimal,YU2010676,lidar2020lecturenotes}, it is sufficient to use a single Wiener process $W(t)$, with real increment $\mathrm{d} W(t)=\xi(t)\mathrm{d}t$. We thus simulated the dynamics of $\ket{\psi_{\xi}(t)}$ linked to Eq.~(\ref{eq:1/f}) employing straightforward Euler-Maruyama technique \cite{Kloeden}. As this approach does not preserve the norm of the state, renormalization after each time step was used. Notice that the dynamics of the reduced density matrix corresponding to this simple instance of stochastic Schr\"odinger equation is equivalent to a quantum master equation in Lindblad form \cite{lidar2020lecturenotes},
\begin{equation}
\dot{\rho}(t)=-i\comm{H_{\text{o}}}{\rho(t)} + \gamma \qty(L\rho(t)L^{\dagger} -\frac{1}{2}\{L^{\dagger}L,\rho(t)\}),
\end{equation}
with $L=L^{\dagger}=\sigma_{z}$, corresponding to a Markovian pure dephasing process.

\section{Dynamical Mode Decomposition (DMD)}
\label{app:DMD}
\begin{figure}
\includegraphics[width=1\linewidth]{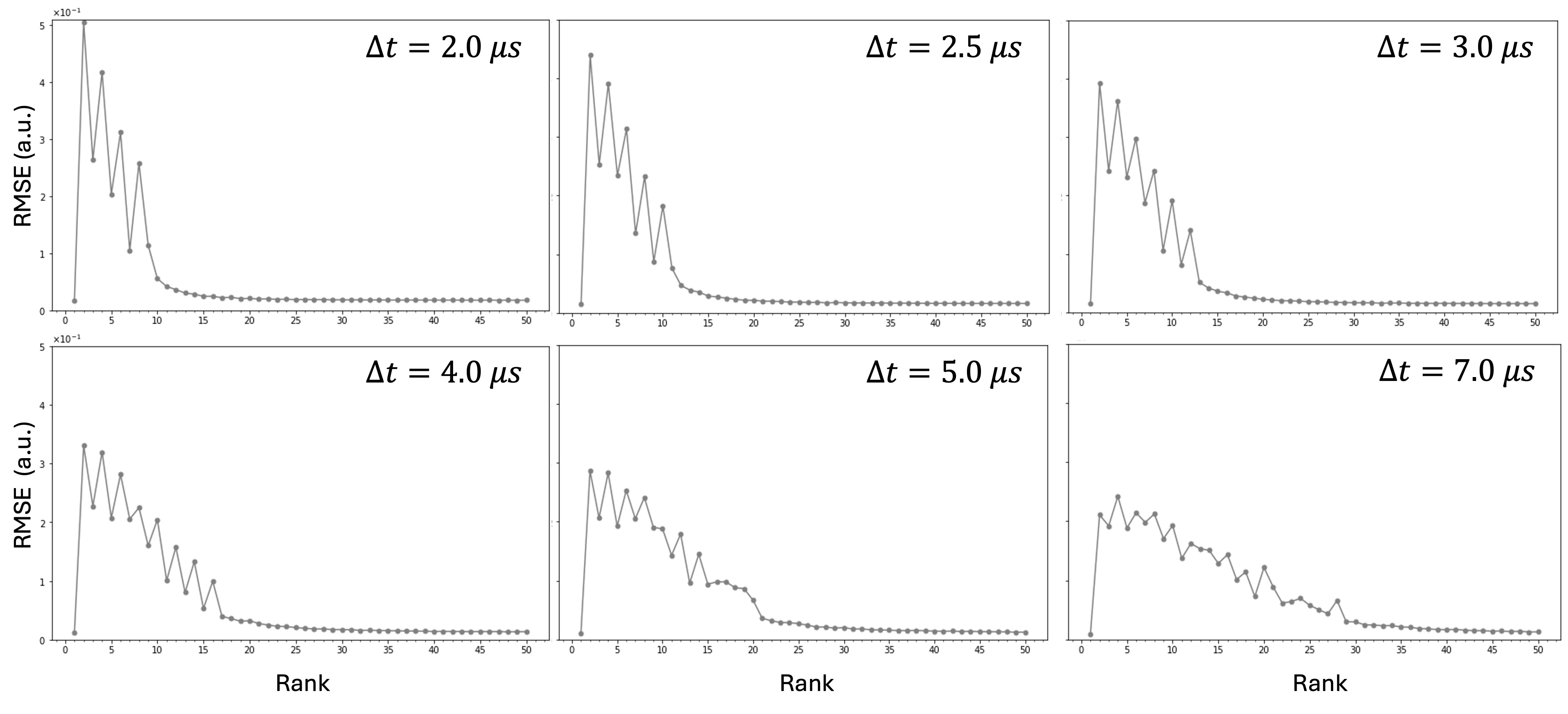} 
\caption{Root-mean-square reconstruction error as a function of DMD rank for six measurement durations $\Delta t = 2.0,\ 2.5,\ 3.0,\ 4.0,\ 5.0,\ 7.0\ \mu\mathrm{s}$. Longer acquisition windows exhibit their RMSE minima at higher ranks, indicating that extended observation times reveal additional coherent behavior arising from the dominance of low-frequency components in the $1/f$ PSD, thereby requiring larger DMD subspaces to capture long-term coherence. 
}
\label{fig:Fig-SI-rmse}
\end{figure} 

DMD is an unsupervised time-series method, originally developed for the study of fluid mechanics, that models a system's evolution by decomposing complex dynamics into spatial-temporal modes~\cite{schmid2010dynamic,Tu2014}. It operates on time-resolved snapshots of the system's state, arranged into two matrices, $X=[x_0, x_1, \ldots, x_{m-2}],\ X'=[x_1, x_2, \ldots, x_{m-1}]$, where each vector $x_k$ is an $n$-dimensional vector representing the system at time $t_k$, with $n$ corresponding to the number of spatial coordinates (or observables). These matrices are approximately related by a linear propagator $\tilde{A}$ with dimensions $n\times n$, such that $X'\approx \tilde{A}X$. Since $\tilde{A}$ is typically high dimensional, its direct diagonalization is computationally expensive. To avoid this, a reduced low-rank operator $A_r$ is constructed. 
This is achieved through the singular value decomposition (SVD) of $X$ given by $X=U{\Sigma}V^{*}$. In this decomposition $U$ and $V$ are orthogonal matrices ($V^*$ denotes the conjugate transpose of $V$), while $\Sigma$ is a diagonal matrix containing the singular values.

To construct a reduced representation, $r$ components are retained, resulting in the reduced matrices $U_r=[u_0,\ldots,vu_{r-1}]$, $V_r=[v_0, \ldots, v_{r-1}]$ and $\Sigma_{r}=\text{diag}(\sigma_{0}, \ldots, \sigma_{r-1})$. The reduced operator $A_r$, which governs the dynamics in the reduced space, is derived using a similarity transformation $A_r= U_r^* \tilde{A}U_r$ and is calculated as $A_r= U^{*}_{r}X^{'}V_{r}\Sigma^{-1}_{r}$~\cite{Tu2014,kutz2016dynamic}.
Following this, the eigenvalues $\lambda_{i}$ and the eigenvectors $W_i$ of the reduced operator $A_r$ are computed. Since $A_r$ is derived from a similarity transformation of $\tilde{A}$, it shares the same eigenvalues $\lambda_{i}$, known as the DMD eigenvalues. The associated DMD modes $\phi_{i}$, are then obtained by projecting the eigenvectors $W_{i}$ onto the measurement space using the relation~\cite{Tu2014,kutz2016dynamic}:
\begin{equation}
    \Phi = X' V_r \Sigma_r^{-1} W.
\end{equation}

Each column $\phi_i$ of $\Phi$ corresponds to a spatial mode associated with frequency defined by the eigenvalue $\lambda_i$.

The eigenvalues $\lambda_i$ are mapped to continuous-time DMD eigenvalues
$\mu_i=\log(\lambda_i)/\Delta t$ by Eq.~(\ref{eq-omega}), and the full DMD reconstruction of the dataset is given by Eq.~(\ref{eq:DMD_reconstruct}) of the main text, the weights $b_i$ in the linear combination are determined from the initial condition $x(t_0)$ using the Moore–Penrose pseudoinverse:
\begin{equation}
b = \Phi^\dagger x(t_0).
\end{equation}

The optimal rank is chosen to minimize the difference between the reconstructed matrix $X_{\mathrm{DMD}}$ and the original input $X_{\mathrm{IN}}$. Figure~\ref{fig:Fig-SI-rmse} shows the root-mean-square error (RMSE), computed according to Eq.~\ref{eq:rmse}), as a function of rank for different measurement window durations $2.0 \leq \Delta t \leq7.0$. The optimal rank is determined when a sufficiently low and stable minimum error is reached. As the duration of the measurement window increases, a higher rank is required to achieve minimal reconstruction error. This trend is expected, since longer observation times reveal the correlations associated with the dominant low-frequency components of the $1/f$ spectral weights, thereby increasing the effective rank of the data matrix.

\begin{equation}
\text{rmse}^n = \sqrt{\frac{1}{mn} \sum_{i=1}^{m} \sum_{j=1}^{n} \left( X_{DMD}^{ij} - X^{ij}_{IN} \right)^2}
\label{eq:rmse}
\end{equation}

\onecolumn

\renewcommand{\thefigure}{S\arabic{figure}}
\renewcommand{\thetable}{S\arabic{table}}
\renewcommand{\thesection}{S\arabic{section}}
\renewcommand{\thesubsection}{S\arabic{section}.\arabic{subsection}}

\setcounter{section}{0}
\setcounter{equation}{0}
\setcounter{figure}{0}

\section*{Supplementary Materials}
\section{Comparison with conventional noise spectrum reconstruction methods}
\label{app:benchmarks}

In this supplementary information, we benchmark our DMD-based reconstruction against
representative conventional approaches for spectral estimation from short,
noisy records. We focus on two classes:
(i) Fourier-based power spectral density (PSD) estimators applied directly
to the observable trajectories, and
(ii) inverse methods based on an analytic forward model linking the
dephasing envelope to the noise PSD. We also comment on harmonic
inversion and Bayesian noise spectroscopy.

Throughout, we use the same simulated data as in the main text:
an ensemble of Ramsey-type trajectories of a qubit evolving under the stochastic Hamiltonian
\begin{equation}
H(t) = H_0 - \frac{1}{2}\,\xi(t)\,\sigma_z,
\qquad
H_0 = \frac{\omega_0}{2}\,\sigma_z,
\label{eq:S_app_hamiltonian}
\end{equation}
with either $1/f$ noise modeled by an ensemble of two-state fluctuators (Methods, Eqs.~(9)-(10)) or white noise generated from a Wiener process with strength $\gamma$ (Methods, Eq.~(15)).
The qubit is initialized in a superposition state, and we analyze the stochastic trajectories of the transverse observable
$\langle\sigma_x(t)\rangle_\xi$ as in the main text.

\subsection{Welch PSD of observable trajectories}

A natural baseline is to compute the PSD of single noisy trajectories using standard Fourier-based methods such as Welch's estimator~\cite{Welch1967}.
Given a discrete time series $x(t_n)$, $n=0,\dots,N-1$, representing a single realization of $\langle\sigma_x(t)\rangle_\xi$, Welch's method proceeds by
(i) segmenting $x(t_n)$ into overlapping windows of length $M$,
(ii) applying a taper (window function) to each segment,
(iii) computing the FFT of each window to form a collection of
periodograms, and (iv) averaging these periodograms to obtain a
variance-reduced PSD estimate.

In our setting, each simulated trajectory is interpreted as a single
experimental realization of the Ramsey signal of the noisy qubit.
We therefore apply Welch's method separately to individual trajectories and to their averages, using a sampling rate
$f_s = 1/\Delta t$ equal to the numerical time step and a fixed window
length $M$ (in the examples shown, $M = 3000$ points).

\begin{figure}[h]
\centering
\includegraphics[scale=0.45]{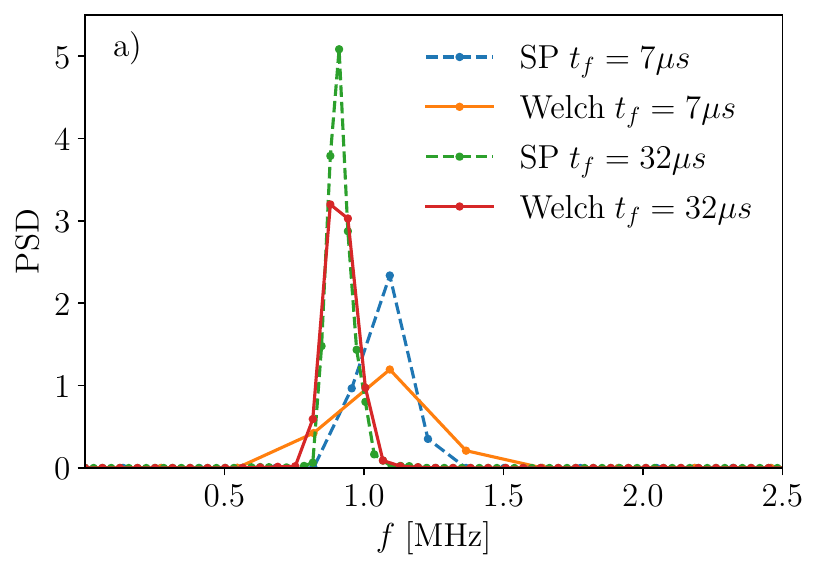}
\label{fig:onea}
~
\includegraphics[scale=0.45]{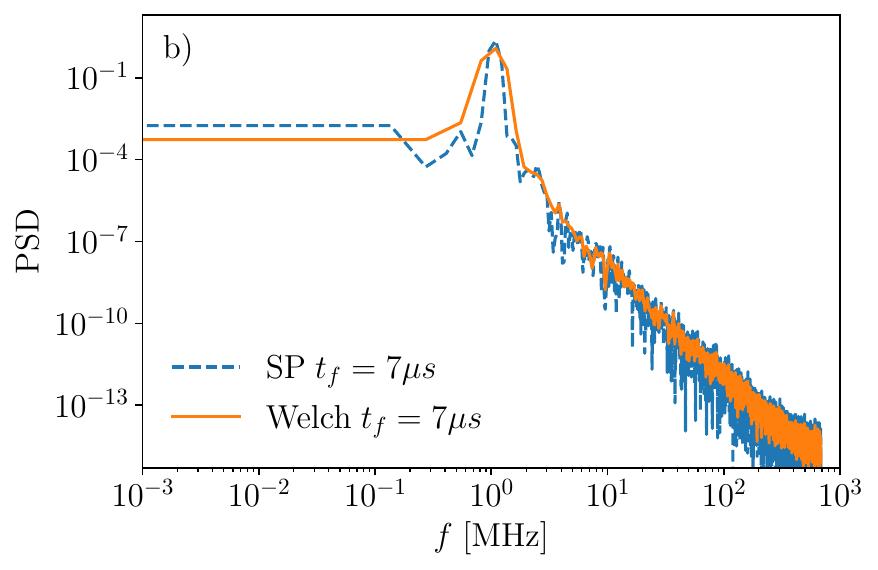}
\label{fig:oneb}
\caption{a) One-sided PSD computed starting from several noisy trajectories, for $1/f$ noise case, where $N_F v^2=0.005$. 
Single periodogram (SP) (dashed lines) are plotted against Welch (solid lines), setting two different final times $t_f=7, 32\mu s$. b) One-sided PSD (Log-Log scale) computed starting from several noisy trajectories,for $1/f$ noise case with $N_F v^2=0.005$. Single periodogram (SP) (dashed blue line) against Welch (solid orange line) are plotted setting two different final times $t_f=7,32\mu s$.}
\label{fig:one}
\end{figure}

\begin{figure}[h]
\centering
\includegraphics[scale=0.45]{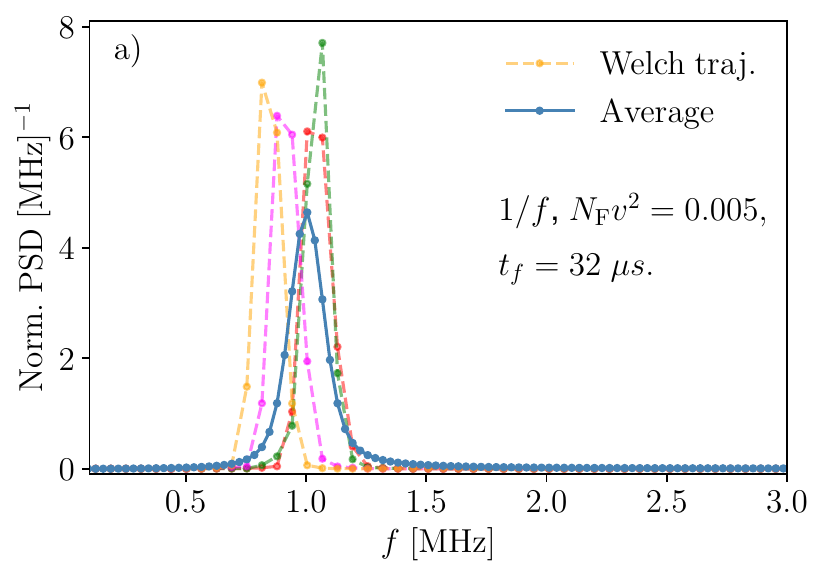}
\label{fig:twoa}
~
\includegraphics[scale=0.45]{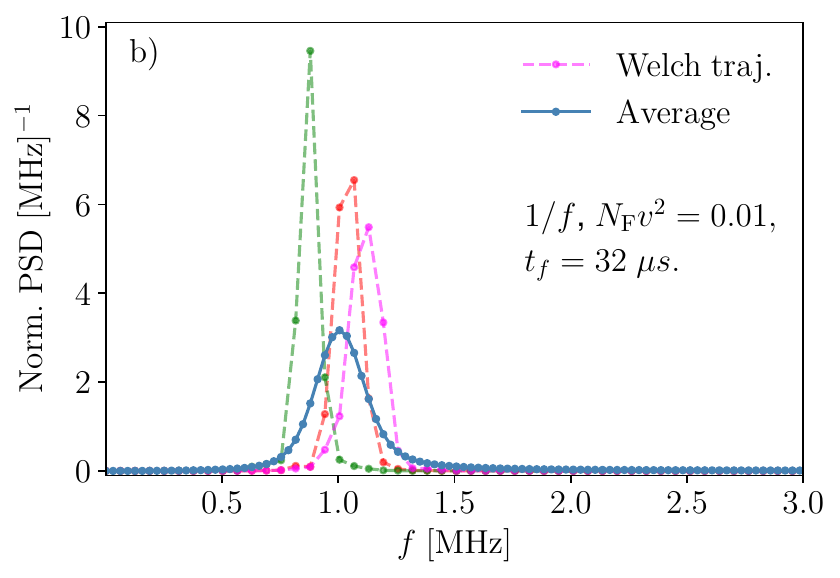}
\label{fig:twob}
\caption{a) Normalized, one-sided PSDs of several noisy trajectories (dashed lines, Welch method), plotted against the PSD of the average signal (solid line, single periodogram), for the $1/f$ noise setting with $N_{\rm F} v^2=0.005$. b) Normalized, one-sided PSDs of several noisy trajectories (dashed lines, Welch method), plotted against the PSD of the average signal (solid line, single periodogram), for the $1/f$ noise setting with $N_{\rm F} v^2=0.01$. } 
\label{fig:two}
\end{figure}

\begin{figure}[h]
\centering
\includegraphics[scale=0.45]{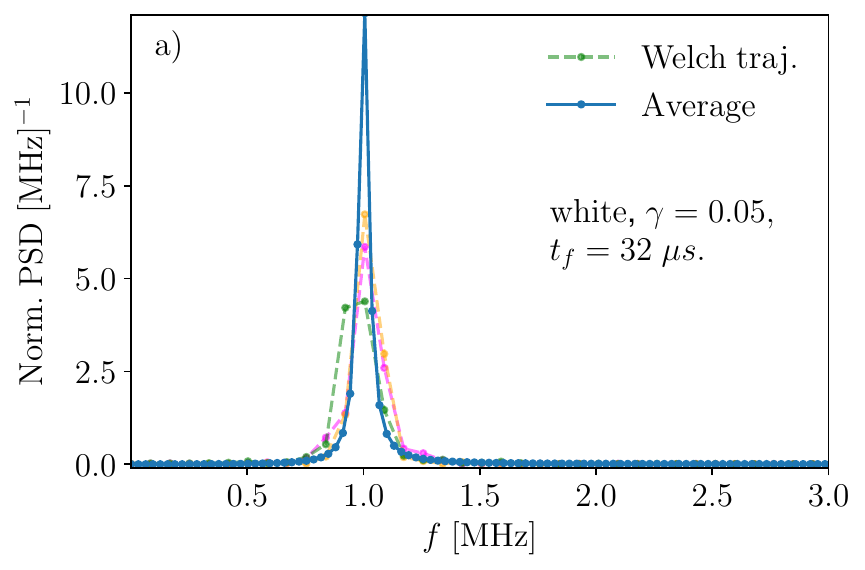}
\label{fig:threea}
~
\includegraphics[scale=0.45]{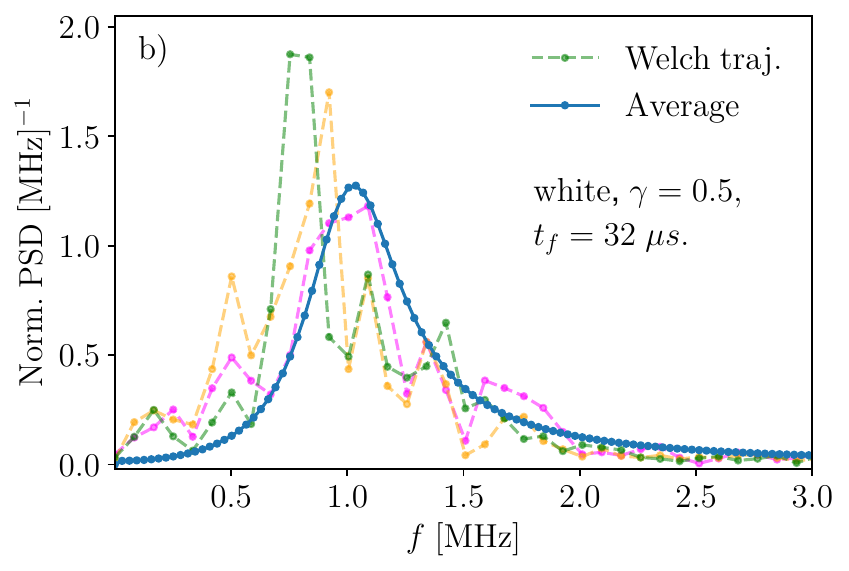}
\label{fig:threeb}
\caption{a) Normalized, one-sided PSDs of several noisy trajectories (dashed lines, Welch method), plotted against the PSD of the average signal (solid line, single periodogram), for the white noise setting with $\gamma=0.05$. b) Normalized, one-sided PSDs of several noisy trajectories (dashed lines, Welch method), plotted against the PSD of the average signal (solid line, single periodogram), for the white noise setting with $\gamma=0.50$.}
\label{fig:three}
\end{figure}

Representative PSDs obtained in this way for the $1/f$ and white-noise cases are shown in the Figs.~\ref{fig:one},~\ref{fig:two}, and~\ref{fig:three}.
In both cases we find:

\begin{itemize}
\item The PSD exhibits a dominant peak centered around a fluctuating frequency $\tilde{f}$, which is of the same order of magnitude of the bare qubit frequency $f_0 = \omega_0 / 2\pi$. However, given that the effect of noise induces a randomization of the qubit frequency (see Eq. \eqref{eq:S_app_hamiltonian}), any different realization of the Ramsey signal provides a different estimate of $\tilde{f}$ (see Fig.~\ref{fig:one}a).
As expected, compared to single periodograms (SP), the Welch method provides smoothed reconstructions of the PSD (see Fig.~\ref{fig:one}b). 
On the contrary, the PSD of the \emph{average} signal, computed by summing over many different trajectories, returns frequency estimates that are consistent with the Ramsey oscillation frequency used in the simulations. Different noise models  (white vs.\ $1/f$) yield PSDs of the average signal whose peak positions agree within the Fourier frequency resolution set by the total simulation time $t_\mathrm{m}$.  

\item Changing the white-noise strength $\gamma$ primarily modifies the overall background level and the width of the peak, but does not substantially shift the peak frequency itself (see Appendix figure comparing PSDs for different $\gamma$). This may not be valid for $1/f$ noise, as the broadening of the peak is the result of a different physical effect \cite{paladino20141} (see Figs.\ref{fig:two} a,~\ref{fig:two}b and Figs.~\ref{fig:three}a,~\ref{fig:three}b).
\end{itemize}

These observations illustrate both the strengths and limitations of
applying Welch's method directly to $\langle\sigma_x(t)\rangle$:

\begin{enumerate}
\item The qubit's characteristic oscillation frequency $f_0$ is reliably extracted from relatively short records only in the limit of small noise amplitude, and this estimate is largely insensitive to the detailed structure of the environmental noise spectrum. On the other hand, the fluctuations of $\tilde{f}$ around $f_{0}$ depend on the details of the noise spectrum, and they cannot be easily modeled starting from the bare knowledge of the single trajectories.

\item By contrast, the PSD of $\langle\sigma_x(t)\rangle$ does not provide a straightforward, quantitative reconstruction of the bath PSD $S(\omega)$ that drives dephasing. The observable PSD is shaped by the system's filter function for the chosen control sequence (here, a Ramsey free-evolution interval) and by the nonlinear mapping from $\xi(t)$ to the coherence envelope. In particular, in the weak-coupling limit, different environmental spectra can yield qualitatively similar PSDs of the observable over a finite bandwidth.

\item Improving the frequency resolution of the Welch estimate requires longer observation times $t_\mathrm{m}$, since the FFT resolution is $\Delta f \sim 1/t_\mathrm{m}$. This is the usual time-frequency
  trade-off for Fourier-based methods and cannot be overcome by averaging over many periodograms alone.
\end{enumerate}

More advanced line-spectral estimators from the signal-processing literature, such as MUSIC, matrix-pencil methods, Prony's method, or atomic-norm approaches, can further refine peak locations and separate
closely spaced spectral lines when the signal is well modeled as a superposition of a small number of sinusoidal components with additive measurement noise. In our Ramsey dephasing setting, however, the primary
objective is different: we aim to reconstruct a broadband \emph{bath} PSD \cite{GardinerZoller,Clerk_2010} from the non-exponential envelope of ensemble-averaged coherence, rather than to resolve discrete spectral lines of a clean underlying signal. For this task, direct application of such line-spectral methods to $\langle\sigma_x(t)\rangle$ does not naturally yield $S(\omega)$.

\subsection{Forward-model inversion and Tikhonov regularization}

A more targeted route to the bath PSD $S(\omega)$ relies on an analytic
forward model that expresses the dephasing envelope in terms of $S(\omega)$.
For classical \emph{Gaussian} dephasing noise $\xi(t)$, the average
transverse coherence
\begin{equation}
\langle\!\langle m_+(t)\rangle\!\rangle
= \langle\!\langle\sigma_x(t)\rangle\!\rangle
+ i \langle\!\langle\sigma_y(t)\rangle\!\rangle
\end{equation}
can be written in the standard filter-function form
\begin{equation}
\langle\!\langle m_+(t)\rangle\!\rangle
= e^{i\phi_0(t)} \left\langle\!\left\langle e^{i\phi(t)}\right\rangle\!\right\rangle
m_+(0),
\label{eq:app_mplus_def}
\end{equation}
where $\phi_0(t) = \omega_0 t$ is the deterministic phase generated by
$H_0$ and $\phi(t) = \int_0^t \xi(t')\,\mathrm{d}t'$ is the stochastic phase
 accumulated from the noise.
Under the Gaussian assumption, $\xi(t)$ is fully characterized by its
two-point correlator or, equivalently, by its PSD $S(\omega)$.
For a Ramsey protocol, one obtains~\cite{paladino20141}
\begin{equation}
\langle\!\langle m_+(t)\rangle\!\rangle
= e^{i\phi_0(t)}\exp\!\left[- \int_0^\infty \frac{\mathrm{d}\omega}{\pi}\,
 \frac{1 - \cos(\omega t)}{\omega^2}\,S(\omega)
\right] m_+(0).
\label{eq:app_filter_function}
\end{equation}

Equation~\eqref{eq:app_filter_function} is exact for classical Gaussian dephasing noise and provides a linear (though ill-conditioned) relation between $\chi(t) \equiv -\log|\langle\!\langle m_+(t)\rangle\!\rangle|$ and
the PSD $S(\omega)$. In our simulations, this framework applies directly to the white-noise case, where the stochastic Schr\"odinger equation is driven by a Wiener process, and the reduced dynamics is equivalent to a Markovian pure-dephasing Lindblad equation (Methods, Eq.~(16)). For the $1/f$ case, the spin-fluctuator model used here can exhibit non-Gaussian statistics, especially when a small number of strongly coupled fluctuators
dominate. In that regime Eq.~\eqref{eq:app_filter_function} should be viewed as an approximate Gaussian surrogate. It should also be noticed that for $1/f$ noise spectrum, the integral on the r.h.s. of Eq.~\eqref{eq:app_filter_function} is divergent.  

To turn Eq.~\eqref{eq:app_filter_function} into an inverse problem, we discretize both time and frequency. Let $t_k$, $k=1,\dots,K$, denote the sampling times of the coherence envelope obtained by averaging the trajectories as in the main text, and let $\{\omega_j\}_{j=1}^L$ be a grid of frequencies with spacings
$\Delta\omega_j$. Removing the known deterministic phase and initial amplitude, we define
\begin{equation}
Y_k \equiv -\log\left|\frac{\langle\!\langle m_+(t_k)\rangle\!\rangle}
{m_+(0)}\right|
\simeq \sum_{j=1}^L Z(t_k,\omega_j) S(\omega_j)\,\Delta\omega_j,
\end{equation}
with kernel
\begin{equation}
Z(t_k,\omega_j) = \frac{1 - \cos(\omega_j t_k)}{\omega_j^2}.
\end{equation}
In matrix notation, this can be written as
\begin{equation}
Y = \tilde Z S,
\label{eq:app_linear_system}
\end{equation}
where $Y\in\mathbb{R}^K$ is a vector with elements $Y_k$, $S\in\mathbb{R}^L$ contains the discretized PSD values $S(\omega_j)$, and $\tilde Z_{kj} = Z(t_k,\omega_j)\,\Delta\omega_j$.

The forward operator $\tilde Z$ is typically severely ill-conditioned~\cite{clason2021regularization,Goulko_2017}, with rapidly decaying singular values. Direct pseudoinversion of
Eq.~\eqref{eq:app_linear_system} therefore yields highly unstable solutions that are dominated by noise and numerical error.
A standard regularization strategy is Tikhonov (ridge) regularization~\cite{Tikhonov_1963,Phillips_1962},
which selects $S$ as the minimizer of the quadratic cost
\begin{equation}
\hat S
= \underset{S}{\arg\min}\;
\left\{
\|\tilde Z S - Y\|_2^2 + \lambda^2 \|S\|_2^2
\right\},
\label{eq:app_tikhonov}
\end{equation}
where $\lambda > 0$ is a regularization parameter.
Interpreted in a Bayesian framework \cite{JARRELL1996133,Mischenko_2012}, Eq.~\eqref{eq:app_tikhonov}
corresponds to a Gaussian likelihood for $Y$ and a zero-mean Gaussian prior $P(S) \propto \exp[-(\lambda^2/2)\|S\|_2^2]$ on the PSD, thus the Tikhonov solution is the maximum a posteriori (MAP) estimate.

\begin{figure}[h]
\centering
\includegraphics[scale=0.45]{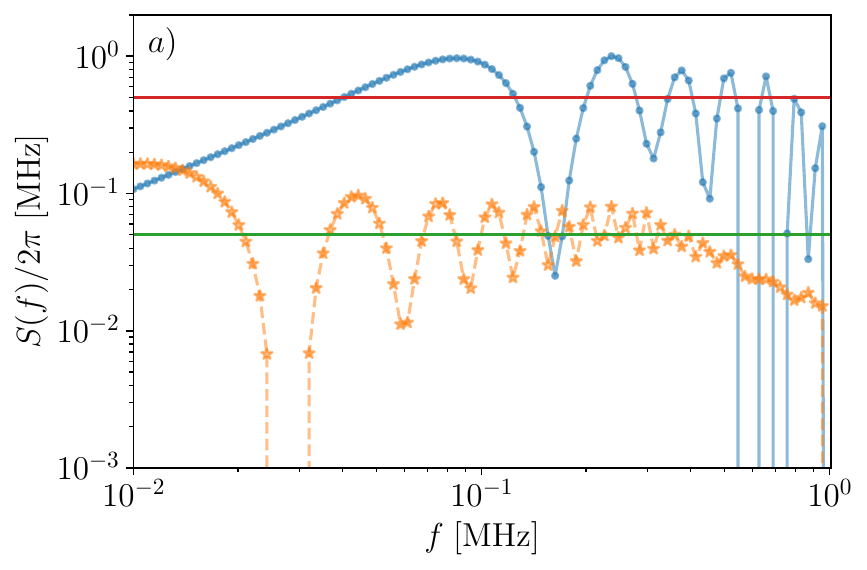}
\label{fig:foura}
~
\includegraphics[scale=0.45]{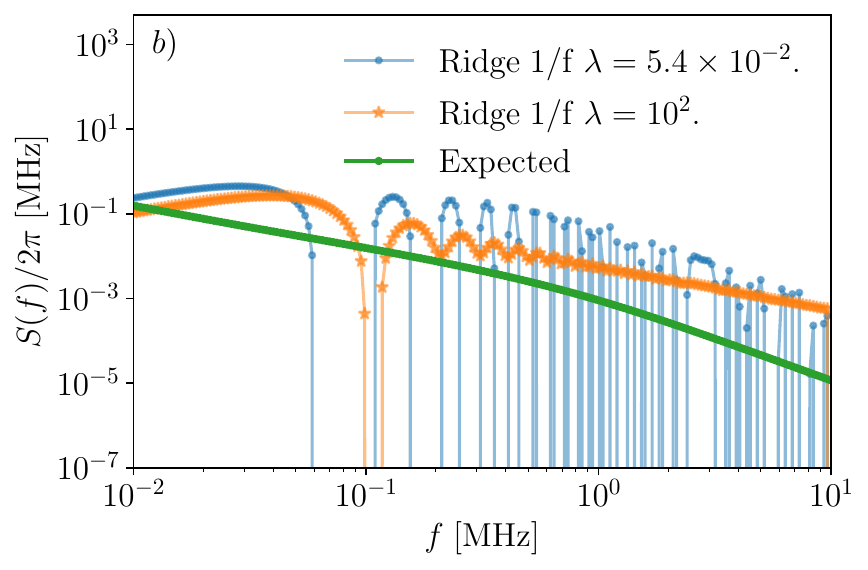}
\label{fig:fourb}
\caption{a) Results of Ridge regression for the white noise case with $\gamma=0.05,0.5$, plotted against the expected value from Eq.\eqref{eq:app_filter_function}.In both cases, the regularization parameters have been set to $\lambda=1.0$. b) Results of Ridge regression for the $1/f$ noise case with $N_{\rm F} v^2=0.005$, plotted against the expected value from the simulation.  The regularization parameters have been set to $\lambda=5.4\cdot 10^{-2},10^{2}$.}
\label{fig:four}
\end{figure}

We applied this scheme to the same simulated coherence envelopes used in the DMD analysis, using logarithmically spaced frequency grids (e.g.\ $L=200$ points for the $1/f$ case and $L=100$ for the white-noise case) and $K=7\cdot 10^3$ time samples. Representative reconstructions are shown in Figs.~\ref{fig:four}a,~\ref{fig:four}b (Tikhonov PSDs). In both the $1/f$ and white-noise examples we observe:

\begin{itemize}
\item For both $1/f$ and white noise, the Tikhonov-inverted spectra reproduce the correct overall scale and gross shape of $S(\omega)$ only over a limited intermediate frequency band. Outside this band, especially at very low and very high frequencies, the reconstructions deviate significantly from the simulated spectra. In contrast, the DMD-based spectra in the main text follow the simulated $S(\omega)$ more faithfully over the same frequency range using the same time window and number of trajectories.

\item The Tikhonov solutions are visibly noisy and depend sensitively on the regularization parameter $\lambda$, reflecting the ill-conditioned singular value spectrum of $\tilde Z$. Small changes in $\lambda$ trade off noise suppression against loss of spectral features. The DMD reconstruction, by avoiding an explicit inversion of Eq.~\eqref{eq:app_linear_system} and working directly with the trajectory ensemble, does not require tuning such a regularization parameter: the spectral weights are fixed by the DMD eigenvalues and the softmax mapping.

\item The kernel $Z(t_k,\omega_j) \propto [1-\cos(\omega_j t_k)]/\omega_j^2$ strongly suppresses the influence of $S(\omega_j)$ at large $\omega_j$ on the data $Y_k$, making the inversion intrinsically insensitive to the detailed high-frequency tail of the spectrum within a finite $t_{\mathrm{m}}$. In the DMD approach, by contrast, high-frequency content that is actually sampled within the time window is encoded directly in rapidly oscillating DMD eigenvalues, so that the same data window can yield a more accurate reconstruction of the accessible part of the spectrum.
\end{itemize}

In the white-noise case, where the Gaussian assumption underlying
Eq.~\eqref{eq:app_filter_function} is satisfied, these limitations
arise purely from ill-conditioning and the restricted time window. In the $1/f$ case, deviations between the Tikhonov reconstruction and the simulated spectrum also reflect the fact that the spin-fluctuator noise can be non-Gaussian for the parameters considered; the Gaussian forward model is then only approximate. The DMD-based reconstruction does not rely on this Gaussian assumption and can be applied in the same way to both noise models considered here.

More sophisticated Bayesian or regularization schemes could be employed to improve these inversions, for example by incorporating smoothness priors\cite{JARRELL1996133}, non-Gaussian likelihoods, or explicit parametric forms for $S(\omega)$ \cite{Ferrie_2018}. However, they invariably require substantial prior modeling
and careful tuning of hyperparameters, and they remain constrained by the same fundamental time-window and conditioning issues illustrated above.

\subsection{Remarks on harmonic inversion and Bayesian noise spectroscopy}

Harmonic inversion and filter-diagonalization methods are designed to extract discrete spectral components (e.g. bound-state energies or normal-mode frequencies) from short-time signals by representing them as superpositions of a small number of damped exponentials.
In favorable cases they can resolve spectral lines with resolution
beyond the naive Fourier limit.
In our Ramsey-dephasing setting, however, the dominant ``signal'' is already a broad dephasing envelope generated by a continuum bath spectrum $S(\omega)$.
Applying harmonic inversion directly to $\langle\sigma_x(t)\rangle$ or its Fourier transform would primarily yield effective poles describing this envelope, rather than a direct reconstruction of the underlying bath PSD.
Relating those poles back to $S(\omega)$ requires additional modeling assumptions and parameterizations of the bath, similar in spirit to the forward-model inversions discussed above.
By contrast, the DMD-based approach does not assume a small number of spectral lines and does not require an explicit parametric form for $S(\omega)$: broadband features of the noise are encoded directly in the distribution of DMD eigenvalues.

Bayesian noise spectroscopy for qubits typically exploits the filter-function formalism Eq.~\eqref{eq:app_filter_function} together with time-dependent control sequences (dynamical decoupling, modulation, etc.) that tailor the filter function to probe specific frequency bands of $S(\omega)$.
These schemes assume analytic expressions for the coherence as a
functional of $S(\omega)$ and combine data from multiple control
sequences to infer the PSD with quantified uncertainty, at the price of introducing explicit prior models and hyperparameters for the spectrum.
In our simulations, by contrast, we consider a simple Ramsey protocol with no time-dependent control beyond a static field, and we work directly with stochastic trajectories without invoking a closed-form expression for the dephasing function in the non-Gaussian regime.
Adapting Bayesian noise spectroscopy in its standard form would
therefore require additional analytic approximations and prior choices that are not needed in our DMD framework.

\leavevmode

These comparisons highlight complementary strengths and weaknesses:

\begin{itemize}
  \item Fourier-based PSD estimators (Welch and related methods) applied to single trajectories reliably recover the qubit's characteristic frequency $f_0$ and give qualitative signatures of broadband versus low-frequency-dominated noise, but they do not directly yield the bath PSD $S(\omega)$.
  Harmonic-inversion methods can sharpen frequency resolution when the signal consists of a small number of spectral lines, but they are not naturally formulated to reconstruct a broadband bath spectrum from a dephasing envelope.

  \item Forward-model inversion with Tikhonov regularization, when an accurate Gaussian dephasing model is available, provides a principled route to $S(\omega)$ from ensemble-averaged coherence data.
  However, the inverse problem is severely ill-conditioned, sensitive to regularization choices and priors, and fundamentally limited by the finite time window and, in non-Gaussian regimes, by model mismatch.
  Bayesian noise-spectroscopy schemes mitigate some of these issues by incorporating stronger priors and tailored control sequences, but at the cost of additional modeling assumptions and experimental overhead.

  \item Our DMD-based approach bypasses the need to specify an explicit forward dephasing model or to design specialized control sequences.
  It operates directly on stochastic trajectories, reinterprets DMD
  modes as statistical weights in realization space, and uses a
  nonlinear softmax mapping to construct a PSD-like spectrum from the DMD eigenvalues.
  As shown in the main text, this DMD-derived
  spectrum reproduces the salient features of the simulated white and $1/f$ noise PSDs using only short time windows and modest numbers of trajectories, while simultaneously providing access to the intrinsic coherence time and enabling constrained extrapolation of the dynamics.
\end{itemize}

These benchmarks clarify how the DMD framework fits
within the broader landscape of noise spectroscopy methods: it trades explicit analytic modeling and control design for a data-driven representation that is well suited to extracting broadband spectral information from limited stochastic trajectory data.

\bibliographystyle{cas-model2-names}

\bibliography{bibliography}


\end{document}